\documentclass[prd,superscriptaddress,onecolumn,showpacs,11pt,msmath,preprintnumbers,showkeys]{revtex4}
\usepackage{wrapfig,rotating}
\usepackage{graphicx,epsfig,color}



\usepackage{graphicx}
\usepackage{dcolumn}
\usepackage{bm}

\def\beq{\begin{equation}}
\def\eeq{\end{equation}}
\def\beeq{\begin{eqnarray}}
\def\eeeq{\end{eqnarray}}

\newcommand \pt {$p_T$}
\def\effs {$\sigma_{\textrm{\scriptsize eff}}$}
\def\2GPD{$_2\mbox{GPD}$}

\def\12{$1\otimes 2$}
\def\22{$2 \otimes 2$}

\def\Qsep{Q_{\mbox{\rm\scriptsize sep}}}
\def\Qsep2{Q^2_{\mbox{\rm\scriptsize sep}}}

\begin{document}

\title{Dynamical approach to MPI in W+dijet and Z+dijet production within the PYTHIA event generator}

\author{B.\ Blok$^{1}$,
 P.\ Gunnellini$^{2}$
\\[2mm] \normalsize $^1$ Department of Physics, Technion -- Israel Institute of Technology,
Haifa, Israel\\
\normalsize $^2$ Deutsches Elektronen-Synchrotron (DESY), Notkestra$\ss$e 85, 22607 Hamburg, Germany}

\begin{abstract}
The new numerical approach that includes \12 mechanism is applied to double parton scattering (DPS) in W+dijet and Z+dijet final state  production in proton-proton collisions at LHC. By using the underlying event (UE) simulation from a \textsc{pythia}~8 tune extracted in hadronic events, we show that, like in the case of a four-jet final state, the inclusion of \12 mechanisms improves the description of experimental data measured at 7 TeV. In addition, predictions for proton-proton collisions at a center-of-mass energy of 14 TeV are shown for DPS- and UE-sensitive observables.
\end{abstract}

  \maketitle
\thispagestyle{empty}

\vfill
\section{\bf Introduction}

Hard {\em Multiple Parton Interactions}\/ (MPI)  play an important role in the description of inelastic proton-proton (pp) collisions at high center-of-mass energies. Starting from the eighties~\cite{TreleaniPaver82,TreleaniPaver85,mufti,dDGLAP} until the last decade~\cite{Treleani,Diehl,DiehlSchafer,Diehl2,Wiedemann,Frankfurt,Frankfurt1,SST,stirling,stirling1,Ryskin,Berger,BDFS1,BDFS2,BDFS3,BDFS4,Gauntnew,kutak,Gauntadd,gieseke1,gieseke2,gieseke3,Sjodmok}, extensive theoretical studies have been performed. Significant progress was made on the simulation of multi-parton collisions in Monte Carlo (MC) event generators \cite{gieseke1,gieseke2,gieseke3,Sjostrand:2007gs,Corke:2011yy,Corke:2010yf,Herwig,Lund}. Multiple parton interactions can serve as a probe for {\em non-perturbative correlations}\/ between partons in the nucleon wave function and are crucial for determining the structure of the Underlying Event (UE) at Large Hadron Collider (LHC) energies. Moreover, they constitute an important background for new physics searches at the LHC. A large number of experimental measurements has been released at the Tevatron \cite{Tevatron1,Tevatron2,Tevatron3} and at the LHC \cite{Atlas,cms1,cms2,Chatrchyan:2013qza,zjj,CMS:2015dna,Gunnellini:2014kwa}, showing a clear evidence of MPI at both soft and hard scales. The latter case is usually referred to as ``Double Parton Scattering'' (DPS), which involves two hard scatterings within the same hadronic collision. The cross section of such an event is generally expressed in terms of $\sigma_{\rm eff}$~\cite{TreleaniPaver82,TreleaniPaver85,mufti,dDGLAP,Treleani,Diehl,DiehlSchafer,Diehl2,Wiedemann,Frankfurt,Frankfurt1,SST,stirling,stirling1,Ryskin,Berger,BDFS1,BDFS2,BDFS3,BDFS4,Gauntnew,Gauntadd,Sjodmok}. In the so-called ``mean-field approximation'', the cross section \effs\ is the effective area which measures the transverse distribution of partons inside the colliding hadrons and their overlap in a collision.

\par
\begin{figure}[htbp]
\begin{center}
\includegraphics[scale=0.65]{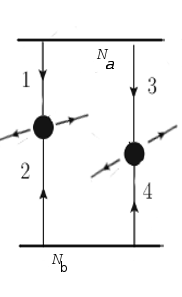}\hspace{3cm}
\includegraphics[scale=0.65]{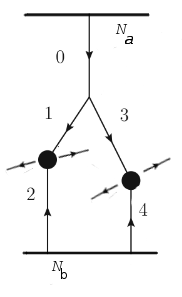}\\
\label{fig1a}
\caption{Sketch of the two considered MPI mechanisms: \22 (left) and \12 (right) mechanism.}
\end{center}
\end{figure}

 Recently, a new approach based in perturbative Quantum Chromodynamics (pQCD) has been developed \cite{BDFS1,BDFS2,BDFS3,BDFS4} for describing MPI. Its main ingredients are listed in the following:
\begin{itemize}
\item the MPI cross sections are expressed through new objects, namely double Generalized Parton distributions (GPD$_2$);
 \item besides the conventional mean-field parton model approach to MPI, represented by the so-called \22 mechanism (see Fig.~\ref{fig1a} left), an additional \12 mechanism (see Fig.~\ref{fig1a} right) is included.  In this mechanism, which can be calculated in pQCD, a parton from one of the nucleons splits at some hard scale and creates two hard partons that participate in MPI. This mechanism leads to a significant transverse-scale dependence of MPI cross sections;
 \item the contribution of the \22 mechanism to GPD$_2$  is calculated in a mean-field approximation with model-independent parameters.
\end{itemize}

\par By using the \textsc{pythia}~8 event generator, this new formalism has been implemented in the simulation of MPI in pp collisions at LHC energies for four-jet final states~\cite{BG}. The new  approach has been developed by including a dynamic \effs\ calculation and has two basic differences from the conventional \textsc{pythia} MPI simulation \cite{Sjostrand:2007gs}. On the one hand, the MPI cross section is calculated in mean-field approach by using the factorization property of GPD$_2$, which connects them to the more conventional one-parton GPD, which have been measured at HERA and parameterized in \cite{Frankfurt,Frankfurt1,BDFS1}. On the other hand, corrections due to \12 mechanisms are implemented by solving nonlinear pQCD evolution equations \cite{BDFS3,BDFS4}. The value of $\sigma_{\rm eff}$ is obtained according to:
\beq\label{mur}
\sigma_{\textrm{\scriptsize eff}}=\frac{\sigma^{(0)}_{\textrm{\scriptsize eff}}}{1+R},
\eeq
where $\sigma^{(0)}_{\textrm{\scriptsize eff}}$ is the effective cross section in the mean-field approach calculated in a model-independent way, and $R(Q_1^2,Q_2^2,Q_0^2)$  is calculated by solving iteratively the nonlinear evolution equation, as explained in detail in \cite{BDFS3,BDFS4}. The two scales $Q_1,Q_2$ are the transverse scales of the two hardest dijet systems produced within the same pp collision, while $Q_0^2=0.5-1$ GeV$^2$ is the scale which divides soft and hard processes in the MPI formalism~\cite{BDFS4}. In such an approach, results obtained for MPI cross sections are model independent, and do not need  additional fit parameters for characterizing the MPI. It has been shown~\cite{BG} that the new formalism implemented in \textsc{pythia}~8 - called in~\cite{BG} and hereafter ``UE Tune Dynamic $\sigma_{\textrm{\scriptsize eff}}$'' - achieves a consistent description of both soft and hard MPI in four-jet final states.

\par The aim of this paper is to extend the MPI approach described in~\cite{BG} for W+dijet and Z+dijet final states. Various predictions are compared to experimental data on W+dijet (hereafter referred to as \textbf{Wjj})~\cite{Atlas,cms1} and inclusive Z production~\cite{zjj} at $\sqrt{s}$~=~7~TeV. The former channel is sensitive to DPS contributions, while observables affected by soft MPI are measured in the latter. Predictions on Z+dijet (hereafter referred to as \textbf{Zjj}) and on inclusive W production at $\sqrt{s}$~=~7~TeV, where experimental data are not yet available, and Zjj, Wjj, inclusive Z, and inclusive W production at $\sqrt{s}$~=~14~TeV are also considered.

\par Unlike a four-jet final state, it has been shown~\cite{Atlas,cms1} that higher-order contributions from single parton scattering (SPS) are crucial for a consistent description of DPS-sensitive observables in the Wjj channel. These are not included in the matrix element simulated by the \textsc{pythia} event generator. Their neglection will lead to incorrect reults, and strong disagreement with experimental data. Other event generators, like \textsc{madgraph}~\cite{madgraph} and \textsc{powheg}~\cite{powheg}, which simulate multileg and higher-order matrix elements, are more suitable for studies of the Wjj and Zjj final states. The \textsc{madgraph} event generator includes both NLO and NNLO real corrections to inclusive W and Z processes but no virtual ones, while \textsc{powheg} generates a full NLO calculation of Wjj and Zjj cross sections with both real and virtual corrections. The matrix elements generated by \textsc{madgraph} and \textsc{powheg} at the parton level are then interfaced to the UE simulation provided by \textsc{pythia}~8.  In particular, the applied UE simulation uses the parameters of the so-called ``UE Tune''~\cite{BG}, extracted from UE data in hadronic events at transverse scales between 2 and 5 GeV. Even though the hard scales are much higher than 5 GeV in inclusive W and Z boson events, our model-independent calculation of $\sigma_{\rm \textrm{\scriptsize eff}}$ in Wjj and Zjj final states shows values which are very similar to $\sigma_{\rm \textrm{\scriptsize eff}}$ obtained in the four-jet channel. This looks however a pure numerical coincidence but motivates the choice of using the ``UE Tune'' also for W and Z boson events. However, it is important to note that the value of the rescaling function $R$ in Eq.~(\ref{mur}) calculated for Wjj and Zjj is very different from the corresponding rescaling function $R$ for four-jet final states. It is only the ratio between $\sigma^{(0)}_{\textrm{\scriptsize eff}}$ and $R$ that is numerically similar between Wjj, Zjj, and four-jet processes.

\par The paper is organized in the following way. In chapter II, the basic formalism and the numerical calculation of \effs\ is described, along with its MC implementation. In chapter III, predictions of observables on Wjj and Zjj final states are compared for different settings of the UE simulation, while chapter IV considers variables on inclusive W and inclusive Z processes. Chapter~V shows predictions of variables in the same previous final states at 14 TeV, while summary and conclusions are given in chapter~VI. Theoretical dependence of \effs\ in Wjj and Zjj as a function of the dijet transverse scale is investigated in appendix~A for 7 and 14 TeV center-of-mass energies, while in appendix~B, results obtained with the \textsc{madgraph} MC event generator are considered.

\section{Basic formalism}
\subsection{Theoretical tools}
\par The approach of the paper is based on the calculation of the MPI cross section by means of the effective cross section \effs. In the case of Wjj or Zjj channels, the MPI cross section can be written as:
\beq
\frac{d\sigma^{\textrm{\tiny Wjj/Zjj}}}{dt_{12}dt_{34}} = \frac{d\sigma^{\textrm{W/Z}}}{dt_{12}} \frac{d\sigma^{\textrm{\tiny dijet}}}{dt_{34}}\times \frac{1}{\sigma_{\textrm{\scriptsize eff}}},
\eeq
where partons 1 and 2 create the gauge boson, and partons 3 and 4 the dijet system. The pQCD calculation leads to the following expression for \effs\ in terms of two-particle GPD:
\begin{eqnarray}
\frac1{\sigma_{\textrm{\scriptsize eff}}} &\equiv&
\int \frac{d^2\vec{\Delta}}{(2\pi)^2}[\mbox{ } _{[2]}GPD_2(x_1,x_3, Q_1^2,Q_2^2;\vec\Delta) _{[2]}GPD_2(x_2,x_4, Q_1^2,Q_2^2; -\vec\Delta)\nonumber\\[10pt]
&+&_{[1]}GPD_2 (x_1,x_3, Q_1^2,Q_2^2;\vec\Delta)
{}_{[2]}G(x_2,x_4, Q_1^2,Q_2^2;-\vec\Delta)\nonumber\\[10pt]
 &+& {}_{[1]}GPD_2(x_2,x_4, Q_1^2,Q_2^2;\vec\Delta){}_{[2]}GPD_2(x_1,x_3, Q_1^2,Q_2^2;-\vec\Delta)].\label{2}
 \end{eqnarray}
The scale $Q_1$of first hard collision is kept fixed to $Q_1=M_W/2$ and $Q_1=M_Z/2$ for, respectively, Wjj and Zjj production. The second and third terms in Eq.~(\ref{2}) correspond to the \12 mechanism, when two partons are generated from the splitting of a parton at a hard scale after evolution. The first term corresponds to the conventional case of two partons evolving from a low scale, namely the \22 mechanism and can be calculated in the mean field approximation \cite{Frankfurt,Frankfurt1,BDFS1}. The momentum $\Delta$ is conjugated to the relative distance between the two participating partons. The full double GPD is a sum of two terms:
\beq
GPD_2(x_1,x_3,Q_1^2,Q_2^2,\Delta)=_{[1]}GPD_2(x_1,x_3,Q_1^2,Q_2^2,\Delta)+_{[2]}GPD_2(x_1,x_3,Q_1^2,Q_2^2,\Delta)\label{3}.
\eeq

\par Here $_{[2]}GPD_2$ corresponds to the part of the GPD$_2$, referring to the occurrence when both partons are evolved from an initial nonperturbative scale. The $_{[1]}GPD_2$ function corresponds to the case when one parton evolves up to some hard scale, at which it then splits into two successive hard partons, each of them participating in turn to the hard dijet event \footnote{We refer the reader to \cite{BDFS1,BDFS2} for the detailed definitions of $_{[1]}GPD_2$ and $_{[2]}GPD_2$ and their connection to nucleon light cone wave functions.}.\\
\par The difference with respect to \cite{BG} is that the partons 1 and 2 are quarks in the case of W and Z boson events (u and $\bar d$ for W and $u\bar u$ or $d\bar d$ for Z production), instead of gluons in the case of four-jet final states. Hence, the GPD in Wjj and Zjj is defined as
\beq
_{[2]}GPD_2(x_1,x_3,Q_1^2,Q_2^2,\Delta)=D_q(x_1,Q_1)D_g(x_3,Q_2)F_{2q}(\Delta,x_1)F_{2g}(\Delta,x_3),
\eeq

where $D(x,Q^2)$ is a conventional parton distribution function (PDF). The use of the mean-field approximation results into:
\beq
_{[2]}GPD_2(x_1,x_3,Q_1^2,Q_2^2,\Delta)=GPD_q(x_1,Q_1^2,\Delta)GPD_g(x_1,Q_1^2,\Delta),\label{slon}
\eeq
and
\beq
GPD_{q,g}(x,Q^2,\Delta)=D_{q,g}(x,Q)F_{2g,2q}(\Delta,x).\label{slon1}
\eeq

\par The numerical analysis of HERA data shows that the gluonic and quark radii of the nucleon are of similar size~\cite{chiral}. Hence, here we neglect the difference between the two-gluon form factors and its quark analogues. By assuming no difference between the initial partons, the structure functions cancel out in the mean-field calculation of \effs. Consequently, the \22 part of the calculation can be done using the two-gluon form factors only. For the two-gluon form factor $F_{2g}$, we use the exponential parametrization described in \cite{Frankfurt1}. In fact, it leads to the same numerical results as the dipole form \cite{Frankfurt}, but it is more convenient for calculations. This parametrization is unambiguously fixed by $J/\Psi$ diffractive charmonium photo/electro production at HERA. The functions $D$ are the conventional nucleon structure functions and $F_{2g}$ can be parameterized as:
\beq
F_{2g}(\Delta,x)=\exp(-B_g(x)\Delta^2/2),
\label{d1}
\eeq
where $B_g$($x$)= $B_0$ + 2$K_Q\cdot\log(x_{0}/x)$, with $x_0\sim 0.0012$, $B_0=4.1$ GeV$^{-2}$ and $K_Q=0.14$ GeV$^{-2}$.
In our implementation the central values of the parameters $B_0$ and $K_Q$ \cite{Frankfurt1} have been used, which are known with an accuracy of $\sim 8\%$. Integrating over $\Delta^2$, we obtain for the part of \effs\ corresponding to the first term in Eq.~(\ref{2}):
\beq
\frac{1}{\sigma^{(0)}_{\textrm{\scriptsize eff}}}=\frac{1}{2\pi}\frac{1}{B_g(x_1)+B_g(x_2)+B_g(x_3)+B_g(x_4)},\label{mura}
\eeq
where $x_{1-4}$ are the longitudinal momentum fractions of the partons participating in the \22 mechanism. This cross section corresponds to the free parton model and is model independent in the sense that its parameters are determined not from a fit to experimental LHC data, but from a fit to single parton GPD. The maximum transversality kinematics for the dijet system, i.e. $4Q^2=x_3x_4s$, have been considered in our approach, being $Q$ the dijet transverse scale, and $x_3,x_4$ the Bjorken fractions of the two jets. Concerning the dependence of \effs\ on the parton scales, the results documented in \cite{BDFS4} show that rescaling factors $R$ in Eq.~(\ref{mura}) are different for different final states, e.g. four-jet, Wjj and Zjj. Hence, we calculated the rescaling factors separately for the two considered channels, Wjj and Zjj. More details on the obtained values of \effs\ as a function of the dijet scale are described in Appendix~B.

\subsection{Monte Carlo implementation and definition of experimental observables}

The standard simulation of MPI implemented in \textsc{pythia}~8~\cite{Corke:2010yf} is considered, but with values of \effs\ calculated by using the QCD-based approach of \cite{BDFS1,BDFS2,BDFS3,BDFS4}, i.e. including \12 processes.

\par The simulation of the MPI in \textsc{pythia} is based on \cite{Corke:2011yy,Corke:2010yf}. The \textsc{pythia} code uses single parton distribution functions, dependent on the impact parameter between the two colliding partons. From a theoretical point of view, these are just GPD$_1$ (see e.g. \cite{DiehlS,radyushkin} for a review). The parameters set in the \textsc{pythia} simulation relative to the transverse parton density are extracted from fits to experimental data on UE, sensitive to the contribution of the MPI. This procedure is closely related to mean-field-based schemes, see e.g.~\cite{BDFS1}.

\par The approach developed in \cite{BG} and used in the present paper combines the standard \textsc{pythia} MPI model with the one of \cite{BDFS1,BDFS2,BDFS3,BDFS4}. We use a single gaussian to model the matter distribution function of the protons. With these settings, the value of $\sigma^{(0)}_{\textrm{\scriptsize eff}}$ would be constant and independent of the scale. In order to implement the \effs\ dependence as a function of the parton momentum fraction $x$ and of the scale, the events where a hard MPI occur, are rescaled according to Eq.~(\ref{mur}). Two types of simulations are considered: one based on the new approach defined in Section~1 and called ``Dynamic $\sigma_{\textrm{\scriptsize eff}}$'', and one which follows the standard \textsc{pythia}~8 approach without any rescaling, called ``UE tune''. \\

\par The UE tune~\cite{BG} has been extracted from fits to UE data in hadronic final states for scales of the leading charged particle in the range 2-5 GeV$^2$ and its parameters are listed in Table~\ref{table1}.

\begin{table}[htbp]
\begin{tabular}{c c} \hline
 \textsc{Pythia}~8 Parameter & Value obtained for the UE tune\\\hline
 MultipartonInteractions:pT0Ref & 2.659\\
 ColourReconnection:range & 3.540\\\hline\hline
 {\effs\ (7 TeV)} (mb) & 29.719\\
 {\effs\ (14 TeV)} (mb) & 32.235\\ \hline
\end{tabular}
\caption{\textsc{Pythia}~8 parameters obtained after the fit to the UE observables. The value of pT0Ref is given at a reference energy of 7 TeV. Values of \effs\ at 7 and 14 TeV are also shown in the table.}
\label{table1}
\end{table}

The first parameter, MultipartonInteractions:pT0Ref, refers to the value of transverse momentum, $p_{\textrm {\scriptsize T}}^0$, defined at $\sqrt{s}=7$ TeV, used for the regularization of the cross section in the infrared limit, according to the formula $1/p_{\textrm {\scriptsize T}}^4\rightarrow 1/(p_{\textrm {\scriptsize T}}^2+p_{\textrm{{\scriptsize T}}}^{0\mbox{ }2})^2$. The second parameter is the probability of colour reconnection among parton strings. The value of \effs\ is found to be around 29.7 mb at 7 TeV and is quite close to the one determined in the mean-field approach \cite{BDFS1,BDFS4}.

Two different sets of observables are studied in final states with W or Z bosons: the first one consists of variables sensitive to hard MPI, the second one includes observables which are mostly influenced by the UE, namely by MPI at moderate scales. Final states with a Z or a W boson are separately investigated at the stable-particle level by using the \textsc{rivet} framework~\cite{Buckley:2010ar}. For the UE study, an inclusive Z or W boson production is required. The Z boson is reconstructed through its muonic decay: two muons with $p_{\textrm {\scriptsize T}}$ $>$ 20 GeV in $|\eta|$ $<$ 2.4 are required with an invariant mass between 81 and 101 GeV$^2$. For the W-boson selection, a final state with one muon with $p_{\textrm {\scriptsize T}}$ $>$ 30 GeV and $|\eta|$ $<$ 2.1 and a missing transverse energy of 30 GeV is required. In the Wjj and Zjj channel, two jets clustered with the anti-k$_{\textrm {\scriptsize T}}$ algorithm with $p_{\textrm {\scriptsize T}}$ $>$ 20 GeV in $|\eta|$ $<$ 2.0 are added to the selection of, respectively, the W and the Z boson. The various final states for which the considered predictions are tested are summarized in Table \ref{table3}.

\begin{table}[htbp]
\begin{tabular}{c c c} \hline
 \textsc{Final state selection} & W-boson & Z-boson\\\hline
UE selection & exactly 1 $\mu$: $p_{\textrm {\scriptsize T}}$ $>$ 30 GeV in $|\eta|$ $<$ 2.1 & 2 $\mu$: $p_{\textrm {\scriptsize T}}$ $>$ 20 GeV in $|\eta|$ $<$ 2.0 \\
& E$^{miss}_T$ $>$ 30 GeV and m$^{W}_{{\scriptsize T}}$ $>$ 50 GeV & m$^{\mu\mu}_{inv}$ in [81,101] GeV$^2$\\\hline
DPS selection &  exactly 1 $\mu$: $p_{\textrm {\scriptsize T}}$ $>$ 30 GeV in $|\eta|$ $<$ 2.1 & 2 $\mu$: $p_{\textrm {\scriptsize T}}$ $>$ 20 GeV in $|\eta$ $<$ 2.0 \\\hline\hline
& E$^{miss}_T$ $>$ 30 GeV and  m$^{W}_{{\scriptsize T}}$ $>$ 50 GeV & m$^{\mu\mu}_{inv}$ in [81,101] GeV$^2$\\\hline
& 2 j: $p_{\textrm {\scriptsize T}}$ $>$ 20 GeV in $|\eta|$ $<$ 2.0 & 2 j: $p_{\textrm {\scriptsize T}}$ $>$ 20 GeV in $|\eta|$ $<$ 2.0\\\hline
\end{tabular}
\caption{Summary of the various selections applied in the W- and Z-boson final states, for studies of UE- and DPS-sensitive observables.}
\label{table3}
\end{table}

\par The following observables are investigated for the study of DPS in Wjj and Zjj final states:
\begin{equation}\label{dels}
\Delta \textrm{S}=\arccos\left(\frac{\vec{p}_T(\textrm{boson})\cdot \vec{p}_T(\textrm{jet}_{1,2})}{|\vec{p}_T(\textrm{boson})|\times |\vec{p}_T(\textrm{jet}_{1,2})|}\right),\\
\end{equation}
\begin{equation}\label{deltarel}
\Delta^{\textrm{\scriptsize rel}}p_{\textrm {\scriptsize T}} = \frac{|\vec{p}_{{\scriptsize T}}^{\textrm{ jet}_1}+\vec{p}_{{\scriptsize T}}^{\textrm{ jet}_2}|}{|\vec{p}_{{\scriptsize T}}^{\textrm{ jet}_1}|+|\vec{p}_{{\scriptsize T}}^{\textrm{ jet}_2}|},\\
\end{equation}
where boson may be the W or Z boson, jet$_{1,2}$ is the jet pair and jet$_1$ (jet$_2$) is the leading (subleading) jet.\\

The study of the UE contribution has been performed through the usual a-la-Rick-Field strategy~\cite{Chatrchyan:2011id}. The direction of the reconstructed boson identifies the direction of the hard scattering and defines different regions in the plane transverse to the beam direction: the ``toward'' region ($|\Delta\phi|$ $<$ 60$^{\circ}$), two transverse regions (60 $<$ $|\Delta\phi|$ $<$ 120$^{\circ}$) and the ``away'' region ($|\Delta\phi|$ $>$ 120$^{\circ}$). Only observables measured in the transverse regions have been considered in this study, since they are the ones which are most affected by the UE contribution. The observables refer to the amount of number of charged particles and of their transverse momentum and are:
\begin{itemize}
\item charged particle multiplicity density ($N_{\textrm{ch}}$);
\item transverse momentum sum density ($\Sigma p_{\textrm {\scriptsize T}}$).
\end{itemize}

Charged particles which contribute to these quantities are selected in each event within a region of pseudorapidity $|\eta|$ $<$ 2.5 with a lower $p_{\textrm {\scriptsize T}}$ cut of 500 MeV. \\

The $x$ and scale dependence of \effs\ has been implemented in the considered predictions by reweighting on an event-by-event basis the MC simulation in presence of a hard MPI ($p_{\textrm {\scriptsize T}}$ $>$ 15 GeV). The $x$ dependence is given by Eq.~(\ref{mura}), where $x_{1,2} $ are taken as the longitudinal momentum fractions of the partons participating in the hardest scattering (W- or Z-boson production), while $x_{3,4} $ refer to the longitudinal momentum fractions of the partons participating in the hardest MPI. The scale dependence is expressed by Eq.~(\ref{mura}), where $R$ takes for $Q_2$ the scale of the hardest MPI. Different values of $Q_0^2$ have been considered in the range between 0.5 and 1 GeV$^2$. Predictions with the following simulation settings are considered for comparison:

\begin{itemize}
\item ``UE Tune''~\cite{BG}: predictions obtained without applying any reweighting of the simulation; this tune uses a constant value of \effs, following the standard \textsc{Pythia} approach, and its parameters have been extracted by fits to the UE measurement in hadronic final states;
\item ``UE Tune $x$-dep'': predictions obtained with the parameters of the UE tune and by applying the $x$ dependence of \effs;
\item ``UE Tune Dynamic \effs'': predictions obtained with the parameters of the UE tune and by applying both $x$ and scale dependence for \effs\ values; two different tunes are shown, corresponding to values of $Q_0^2$ equal to 0.5 and 1 GeV$^2$. As it was done in four-jet final states~\cite{BG}, the dependence of the cross section on the \pt\ of the outgoing partons is assumed to be the same  in \22 and \12 production mechanisms. This might affect differential distributions as a function of the jet balance, i.e. $\Delta^{\textrm{\scriptsize rel}}p_{\textrm {\scriptsize T}}$.
\end{itemize}

\par This approach is implemented for various Monte Carlo event generators which use different matrix-element calculations:
\begin{itemize}
\item \textsc{pythia}~8~\cite{Sjostrand:2007gs}, which implements a 2$\rightarrow$2 LO ME ($q\bar{q}$ $\rightarrow$ Z and $q\bar{q'}$ $\rightarrow$ W for, respectively, Z- and W-boson production), where additional hard partons in the final state are generated through the parton-shower simulation in a leading-log approximation;
\item \textsc{powheg}~\cite{powheg} interfaced to \textsc{pythia}~8, which implements a 2$\rightarrow$4 NLO ME;
\item \textsc{madgraph}~\cite{madgraph} interfaced to \textsc{pythia}~8, which implements a 2$\rightarrow$4 LO ME, where up to four partons in addition to the Z- or the W-boson are simulated within the ME calculation.
\end{itemize}

\begin{figure}[htbp]
\begin{center}
\includegraphics[scale=0.67]{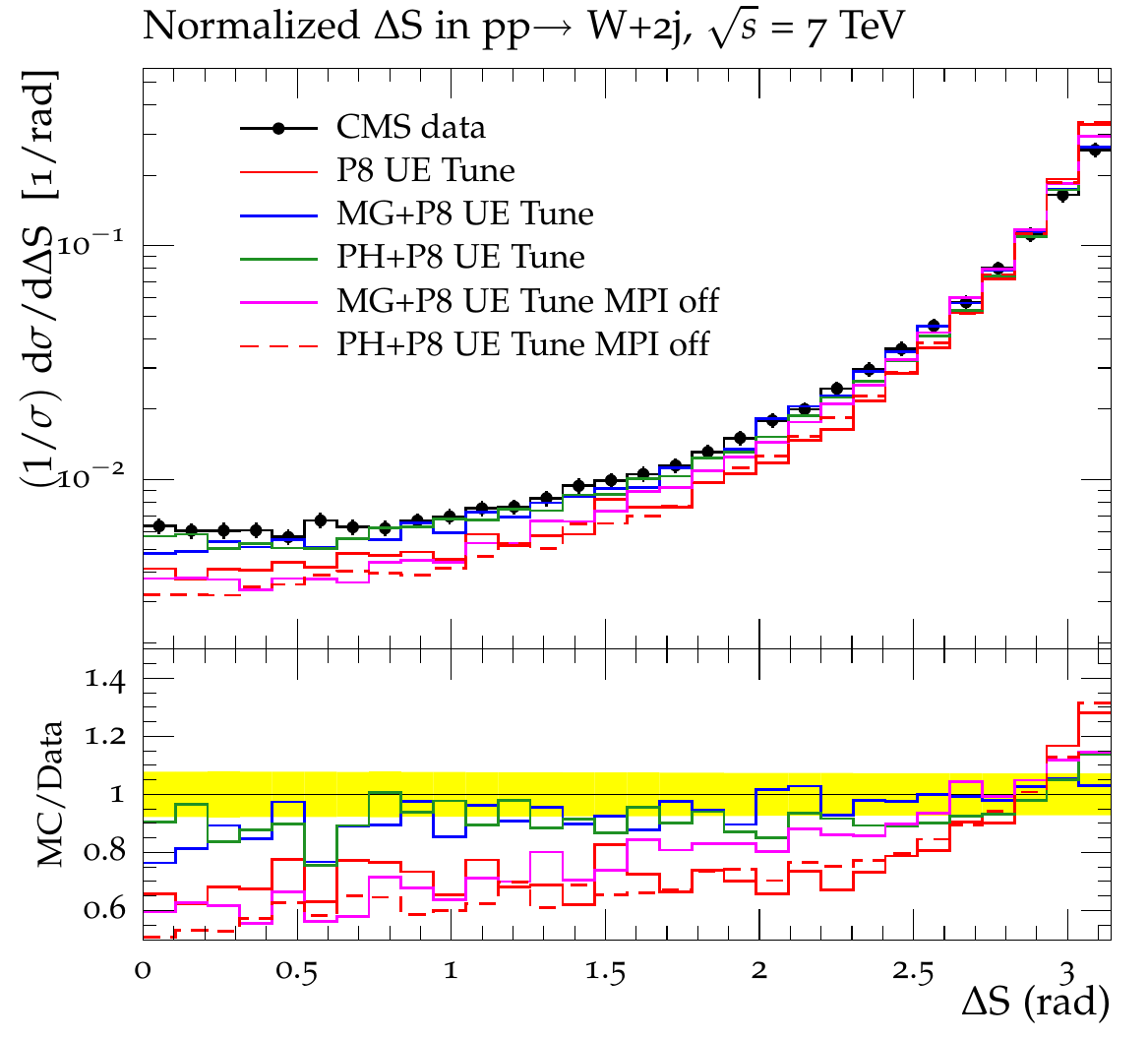}
\includegraphics[scale=0.67]{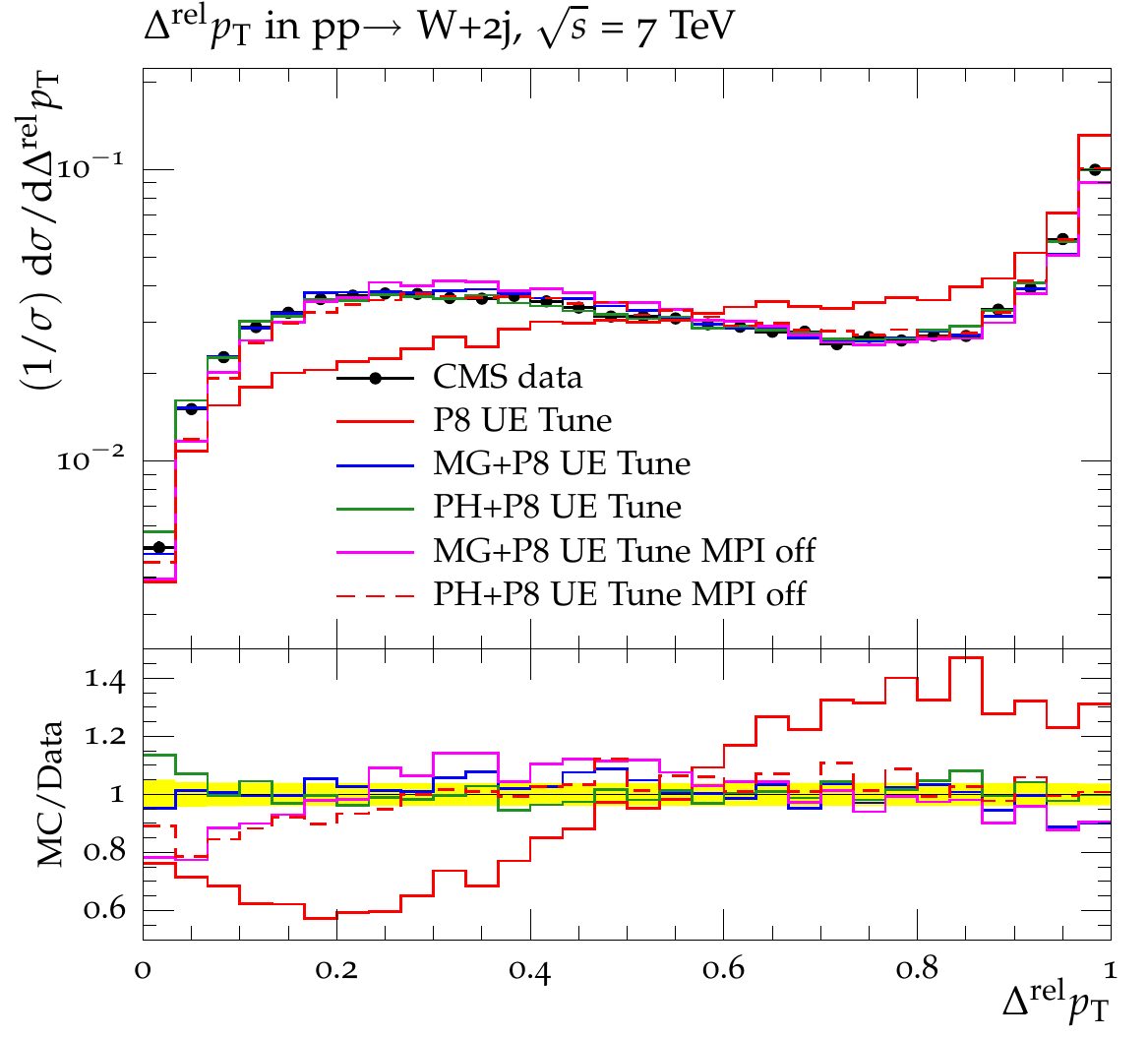}\\
\caption{CMS data~\cite{cms1} at 7 TeV for the normalized distributions of the correlation observables $\Delta$S (\textit{left}) and $\Delta^{\textrm{\scriptsize rel}}p_{\textrm {\scriptsize T}}$ (\textit{right}) in the W+dijet channel, compared to predictions generated with \textsc{pythia}~8 UE Tune, \textsc{madgraph} and \textsc{powheg} interfaced to \textsc{pythia}~8 UE Tune. Predictions obtained with \textsc{madgraph} and \textsc{powheg} interfaced to \textsc{pythia}~8 UE Tune, without the simulation of the MPI are also compared to the measurement. The ratios of these predictions to the data are shown in the lower panels.}
\label{fig1}
\end{center}
\end{figure}

The \textsc{pythia}~8 sample uses the CTEQ6L1~\cite{Pumplin:2002vw} PDF set, while the \textsc{powheg} sample has been generated with the CT10NLO~\cite{Lai:2010vv} PDF set. For the \textsc{madgraph} sample, the CTEQ6L1~\cite{Pumplin:2002vw} PDF set has been used and the matching and merging scale between matrix element (ME) and parton shower (PS) have been set to, respectively, 10  and 20 GeV in the MLM formalism~\cite{Alwall:2007fs}. Predictions obtained with the considered event generators have been compared to the $\Delta$S and $\Delta^{\textrm{\scriptsize rel}}p_{\textrm {\scriptsize T}}$ observables, measured at 7 TeV by the CMS experiment in the Wjj channel~\cite{cms1} and they are shown in Fig.~\ref{fig1}. Predictions obtained with \textsc{powheg} and \textsc{madgraph} interfaced to \textsc{pythia}~8 UE Tune are able to follow the shape of the measured points, while \textsc{pythia}~8 does not describe at the same level of the agreement. In particular, higher-order contributions fill the region of the phase space which is most sensitive to the DPS signal. This effect, already observed in \cite{cms1}, is a clear indication of the need of higher-order matrix elements to give a reasonable description of DPS-sensitive observables. Predictions obtained with \textsc{powheg} and \textsc{madgraph} without the simulation of MPI, also shown in Figure~\ref{fig1}, are not able to follow the trend of the measured $\Delta$S and $\Delta^{\textrm{\scriptsize rel}}p_{\textrm {\scriptsize T}}$. In particular, they underestimate the region of $\Delta$S $<$ 2 by about 50--70\% and the region of $\Delta^{\textrm{\scriptsize rel}}p_{\textrm {\scriptsize T}}$ $<$ 0.15 by about 10--20\%. These are the regions of the phase space where a signal from hard MPI is expected to contribute. The large discrepancy observed between data and predictions without the simulation of MPI clearly indicates the need of MPI contributions in the current models for a good description of DPS-sensitive observables.

\par In the following sections, results are shown by using the \textsc{powheg} event generator interfaced to \textsc{pythia}~8, which consistently includes both real and virtual NLO corrections for hard Wjj and Zjj processes. Comparisons with simulations obtained with \textsc{madgraph}, which was used for experimental extractions of \effs\ by the CMS collaboration~\cite{cms1}, are documented in Appendix~A, while predictions with \textsc{pythia}~8 standalone are dropped from the discussion.

\section{DPS-sensitive observables in W+dijet and Z+dijet final states}

\begin{figure}[htbp]
\begin{center}
\includegraphics[scale=0.67]{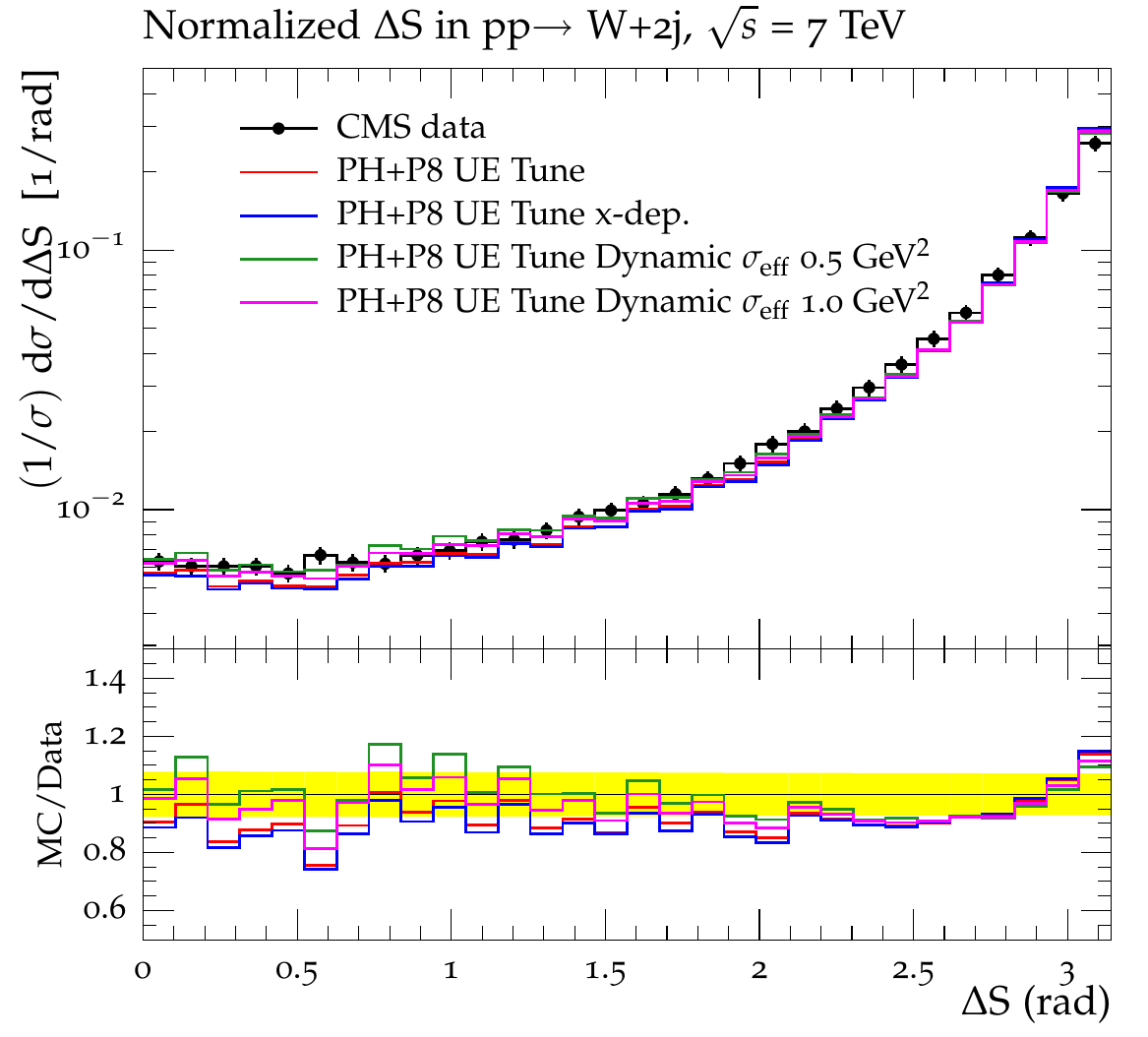}
\includegraphics[scale=0.67]{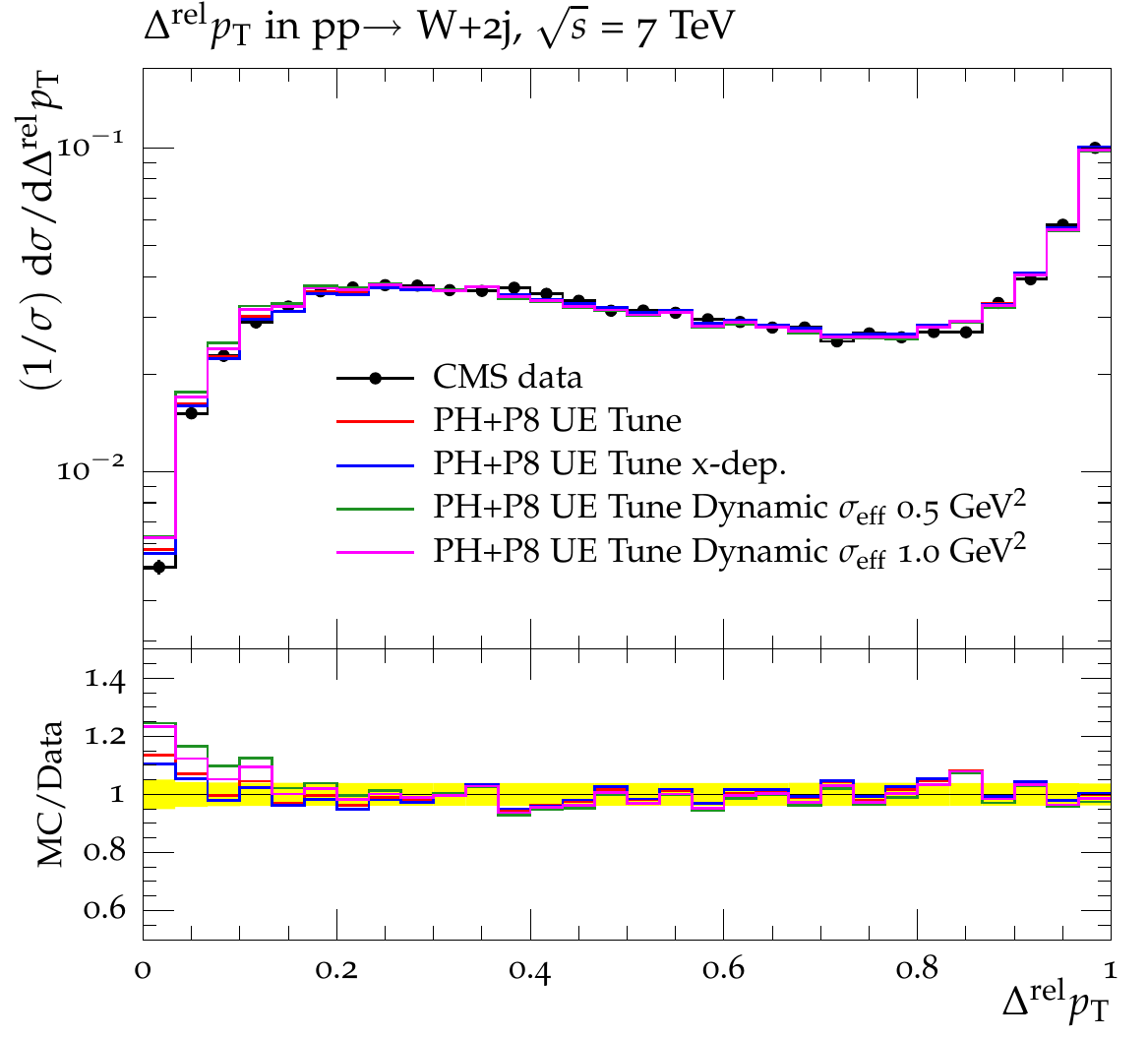}\\
\caption{CMS data at 7 TeV for the normalized distributions of the correlation observables $\Delta$S (\textit{left}) and $\Delta^{\textrm{\scriptsize rel}}p_{\textrm {\scriptsize T}}$ (\textit{right}) in the W+dijet channel, compared to predictions of \textsc{powheg} interfaced to \textsc{pythia}~8 UE Tune with different \effs\ dependence applied: no reweighting applied (red line), $x$-dependent \effs\ values (blue line), $x$- and scale-dependent \effs\ values with $Q^2_0$ $=$ 0.5 GeV$^2$ (green line) and $x$- and scale-dependent \effs\ values with $Q^2_0$ $=$ 1 GeV$^2$ (pink line). Also shown are the ratios of these tunes to the data.}
\label{fig3}
\end{center}
\end{figure}

In this section, comparisons of various predictions for DPS-sensitive observables in Wjj and Zjj are shown at 7 TeV. In Fig.~\ref{fig3}, comparisons to data measured by the CMS experiment~\cite{cms1,cms2,Chatrchyan:2013qza} in the Wjj channel at 7 TeV are considered. They refer to the normalized distributions of the correlation observables $\Delta$S (\textit{left}) and $\Delta^{\textrm{\scriptsize rel}}p_{\textrm {\scriptsize T}}$ (\textit{right}). Predictions of \textsc{powheg} interfaced to \textsc{pythia}~8 UE Tune are considered with different \effs\ dependence applied: no reweighting, only $x$-dependent \effs\ values calculated in mean-field approach, $x$- and scale-dependent \effs\ values with $Q^2_0$ $=$ 0.5 GeV$^2$ and $Q^2_0$ $=$ 1 GeV$^2$.

\par The inclusion of contributions from \12 production mechanisms in the predictions with dynamic \effs\ values improves the agreement with the measurement for the $\Delta$S observable. They follow the decreasing shape of the observable better than the predictions obtained with the UE Tune without any rescaling and the UE Tune with only x-dependence applied. The $\Delta^{\textrm{\scriptsize rel}}p_{\textrm {\scriptsize T}}$ observable is in good agreement with every prediction, except at values $\Delta^{\textrm{\scriptsize rel}}p_{\textrm {\scriptsize T}}$ $<$ 0.15, where the curves are above the data by about 10-25\%. This might be due to the fact that in our approach we assume the $\Delta^{\textrm{\scriptsize rel}}p_{\textrm {\scriptsize T}}$ dependence of the cross section to be the same in \22 and \12 production mechanisms. This does not need to be true \cite{BDFS1,BDFS2,BDFS3,BDFS4}. It would be interesting to study if indeed different assumptions on the differential \12 and \22 cross sections bring to significant difference and play a role at low values of $\Delta^{\textrm{\scriptsize rel}}p_{\textrm {\scriptsize T}}$. However, this demands additional analytical and numerical work, and will be done elsewhere \cite{BG1}. Also, it would be interesting to study the agreement of the considered predictions for a different scale of the dijet system, different from 20 GeV as in this case.

\begin{figure}[htbp]
\begin{center}
\includegraphics[scale=0.67]{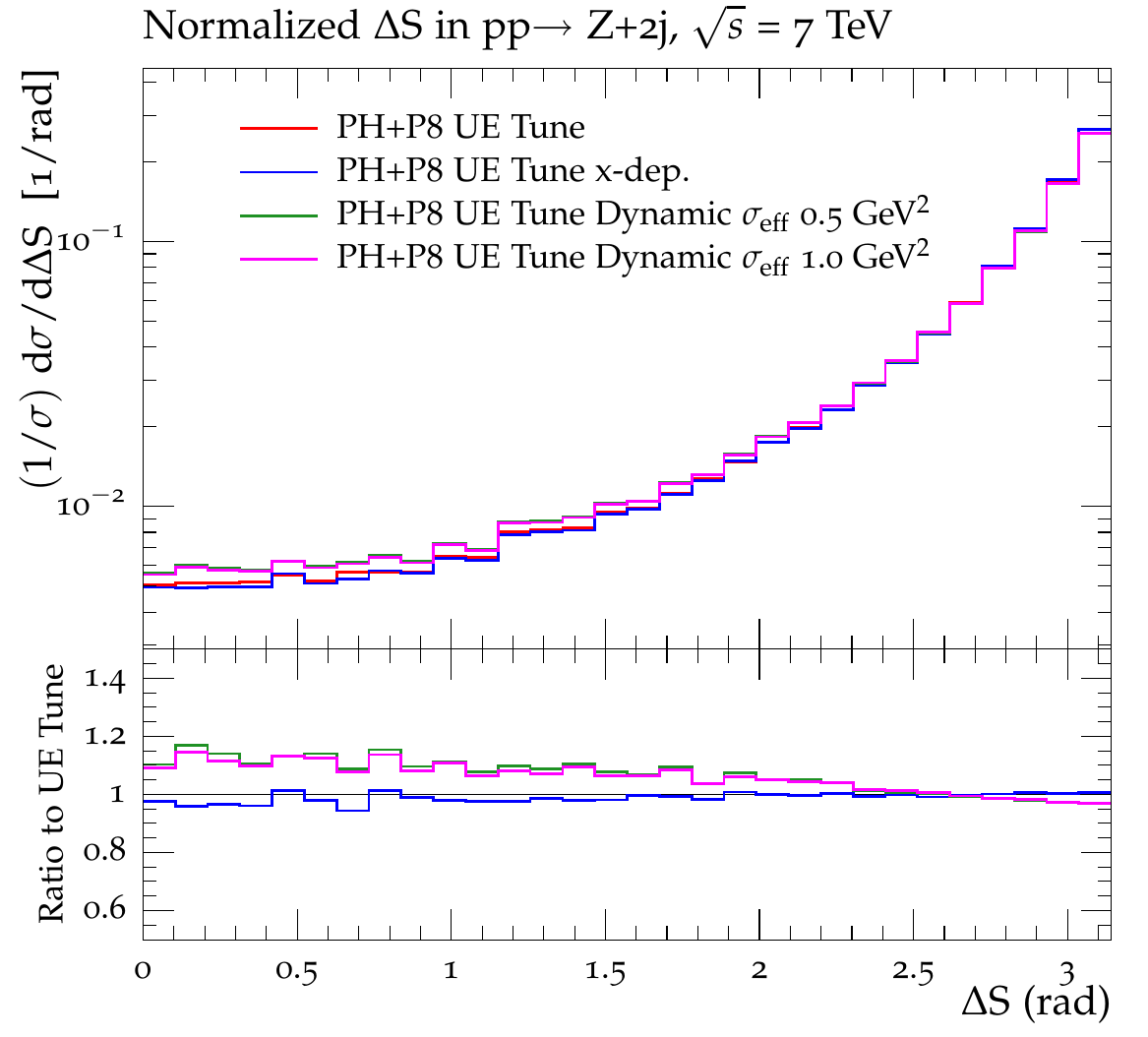}
\includegraphics[scale=0.67]{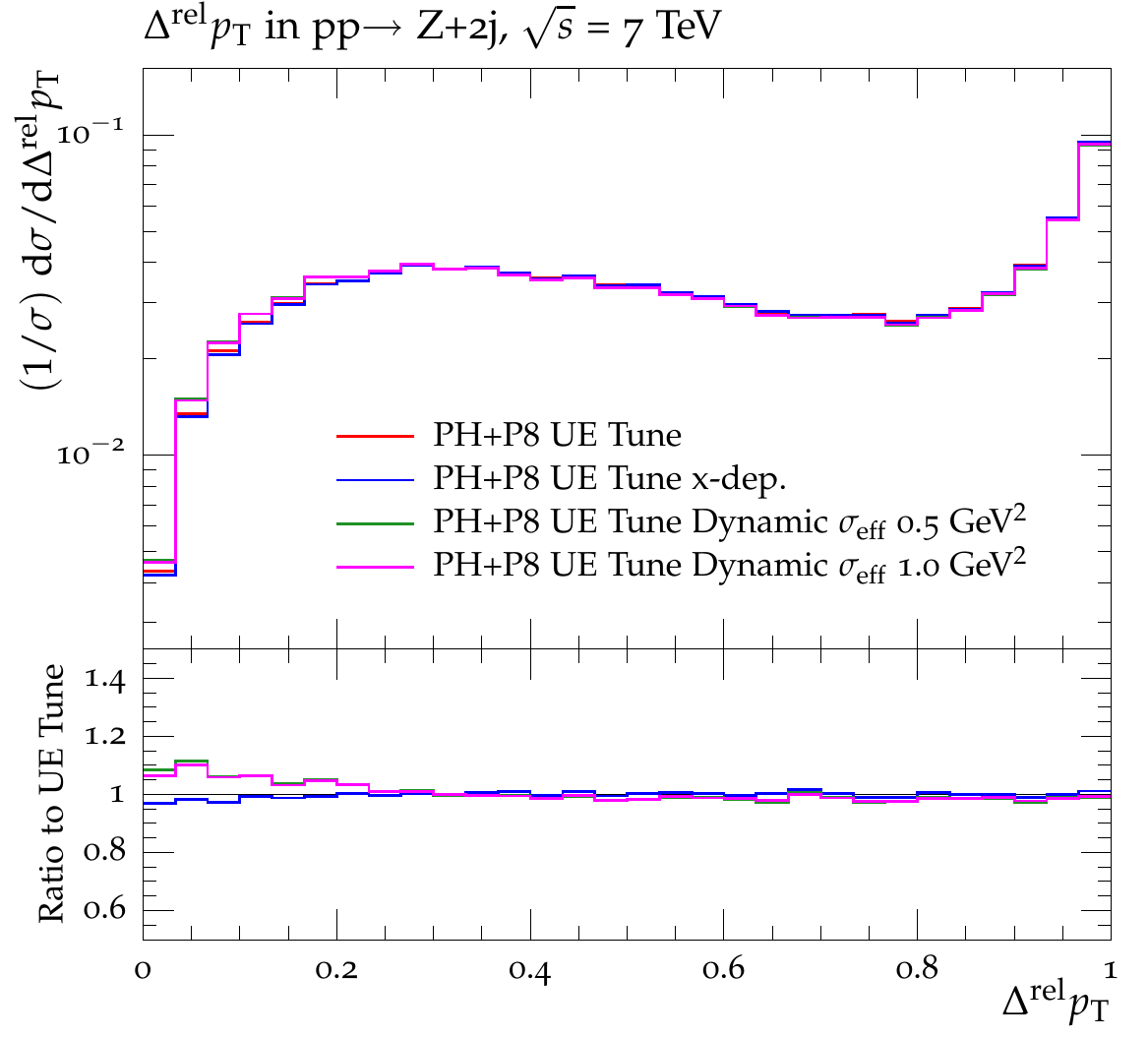}\\
\caption{Predictions at 7 TeV for the normalized distributions of the correlation observables $\Delta$S (\textit{left}) and $\Delta^{\textrm{\scriptsize rel}}p_{\textrm {\scriptsize T}}$ (\textit{right}) in the Z+dijet channel, of simulations performed with \textsc{powheg} interfaced to \textsc{pythia}~8 UE Tune with different \effs\ dependence applied: no \effs\ reweighting applied (red line), $x$-dependent \effs\ values (blue line), $x$- and scale-dependent \effs\ values with $Q^2_0$ $=$ 0.5 GeV$^2$ (green line) and $x$- and scale-dependent \effs\ values with $Q^2_0$ $=$ 1 GeV$^2$ (pink line). Also shown are the ratios of each curve to the predictions of the UE Tune.}
\label{fig9}
\end{center}
\end{figure}

In Fig.~\ref{fig9}, predictions obtained with \textsc{powheg} interfaced to \textsc{pythia}~8 UE Tune are shown for the normalized distributions of the correlation observables $\Delta$S and $\Delta^{\textrm{\scriptsize rel}}p_{\textrm {\scriptsize T}}$ in the Zjj channel. Various simulation settings are considered: no \effs\ reweighting applied-UE tune, $x$-dep tune  values $x$- and scale-dependent xscale tune, with $Q^2_0$ $=$ 0.5 GeV$^2$ and $x$- and scale-dependent \effs\ values with $Q^2_0$ $=$ 1 GeV$^2$. Data points for these observables are not yet measured. Differences among the predictions are of the order of 10-15\% for $\Delta$S $<$ 2 and $\Delta^{\textrm{\scriptsize rel}}p_{\textrm {\scriptsize T}}$ $<$ 0.2, which are the regions of the phase space where DPS signals are expected to contribute.

\section{UE observables in inclusive W and Z boson events}

\begin{figure}[htbp]
\begin{center}
\includegraphics[scale=0.6]{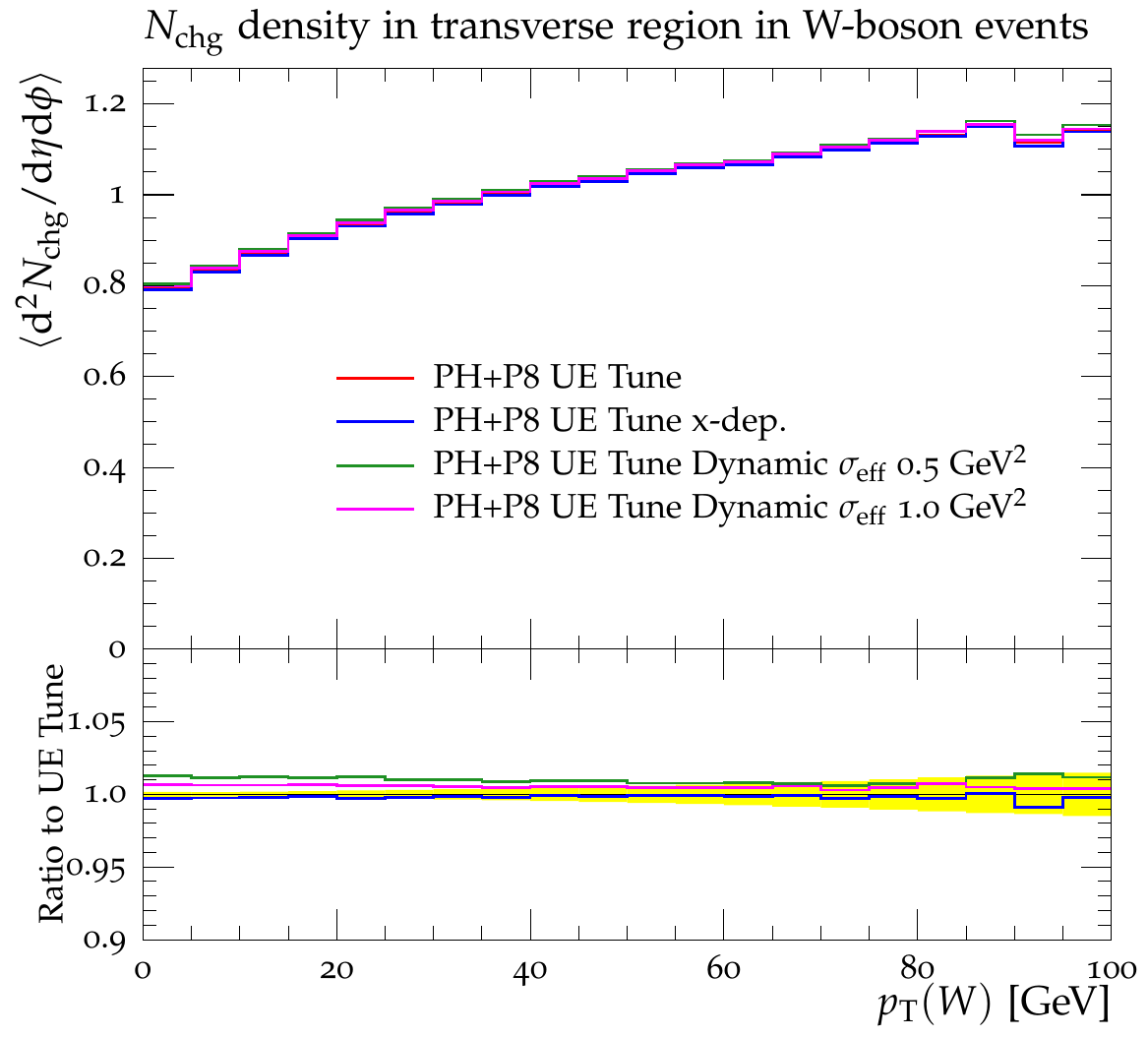}
\includegraphics[scale=0.6]{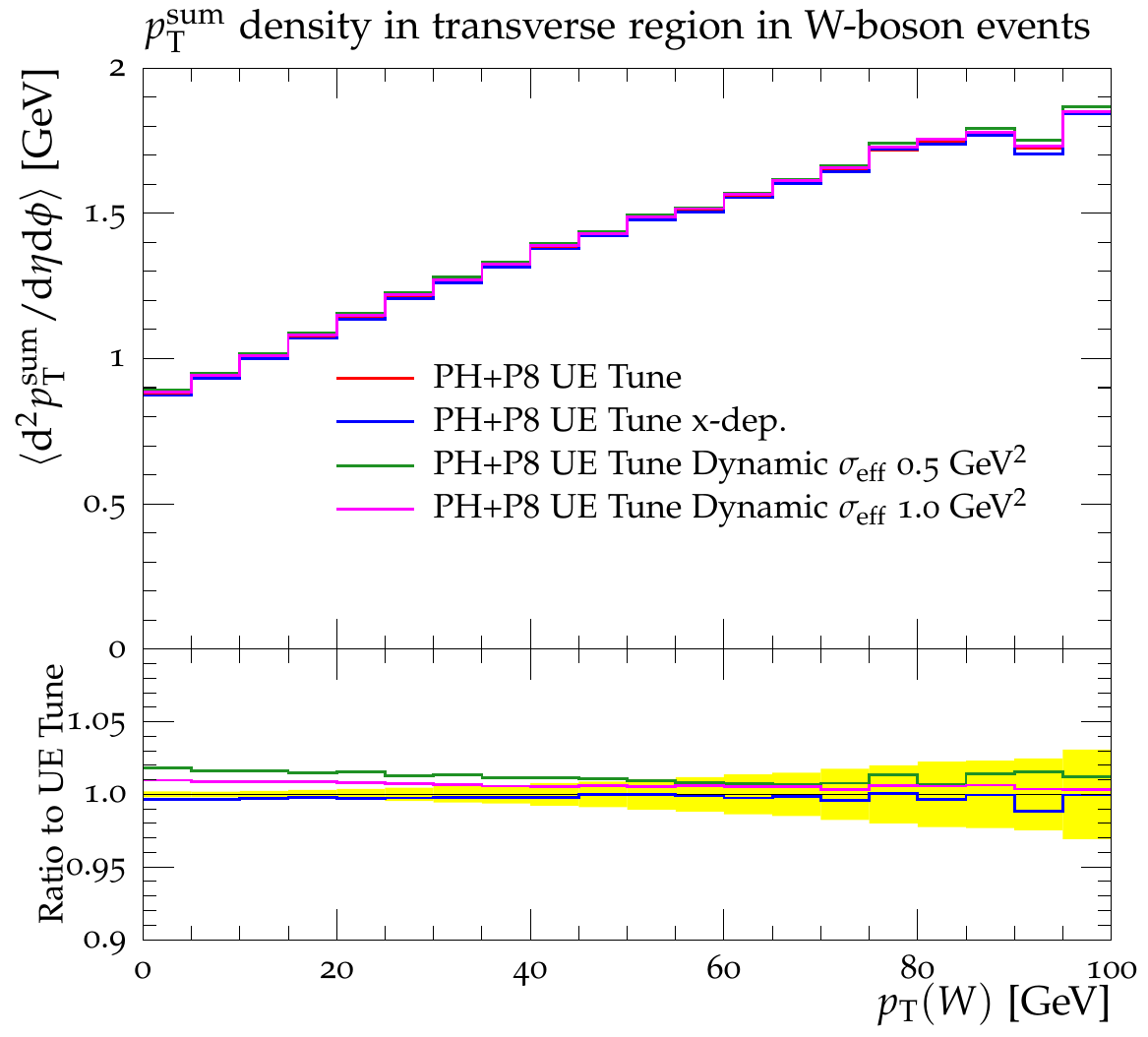}\\
\caption{Predictions for the (\textit{left}) charged-particle and (\textit{right}) $p_{\textrm {\scriptsize T}}$ sum densities in the transverse regions as defined by the W-boson in proton-proton collisions at 7 TeV. Simulations obtained with \textsc{powheg} interfaced to \textsc{pythia}~8 UE Tune are considered with different \effs\ dependence applied: no reweighting applied (red line), $x$-dependent \effs\ values (blue line), $x$- and scale-dependent \effs\ values with $Q^2_0$ $=$ 0.5 GeV$^2$ (green line) and $x$- and scale-dependent \effs\ values with $Q^2_0$ $=$ 1 GeV$^2$ (pink line). Also shown are the ratios of these tunes to the predictions of the UE Tune.}
\label{fig6}
\end{center}
\end{figure}

Predictions obtained with the considered tunes are also tested for UE observables in inclusive W and Z boson events. This kind of events are sensitive to MPI at moderate scales. In Fig.~\ref{fig6}, predictions on charged-particle multiplicity and $p_{\textrm {\scriptsize T}}$ sum densities are shown for inclusive W events in the transverse region as a function of the W-boson $p_{\textrm {\scriptsize T}}$. Curves obtained with \textsc{powheg} interfaced to \textsc{pythia}~8 UE Tune and implementing different \effs\ dependence differ  less than 2\% from each other. This effect is very similar to the one observed in hadronic events, documented in \cite{BG}.

\par
\begin{figure}[htbp]
\begin{center}
\includegraphics[scale=0.6]{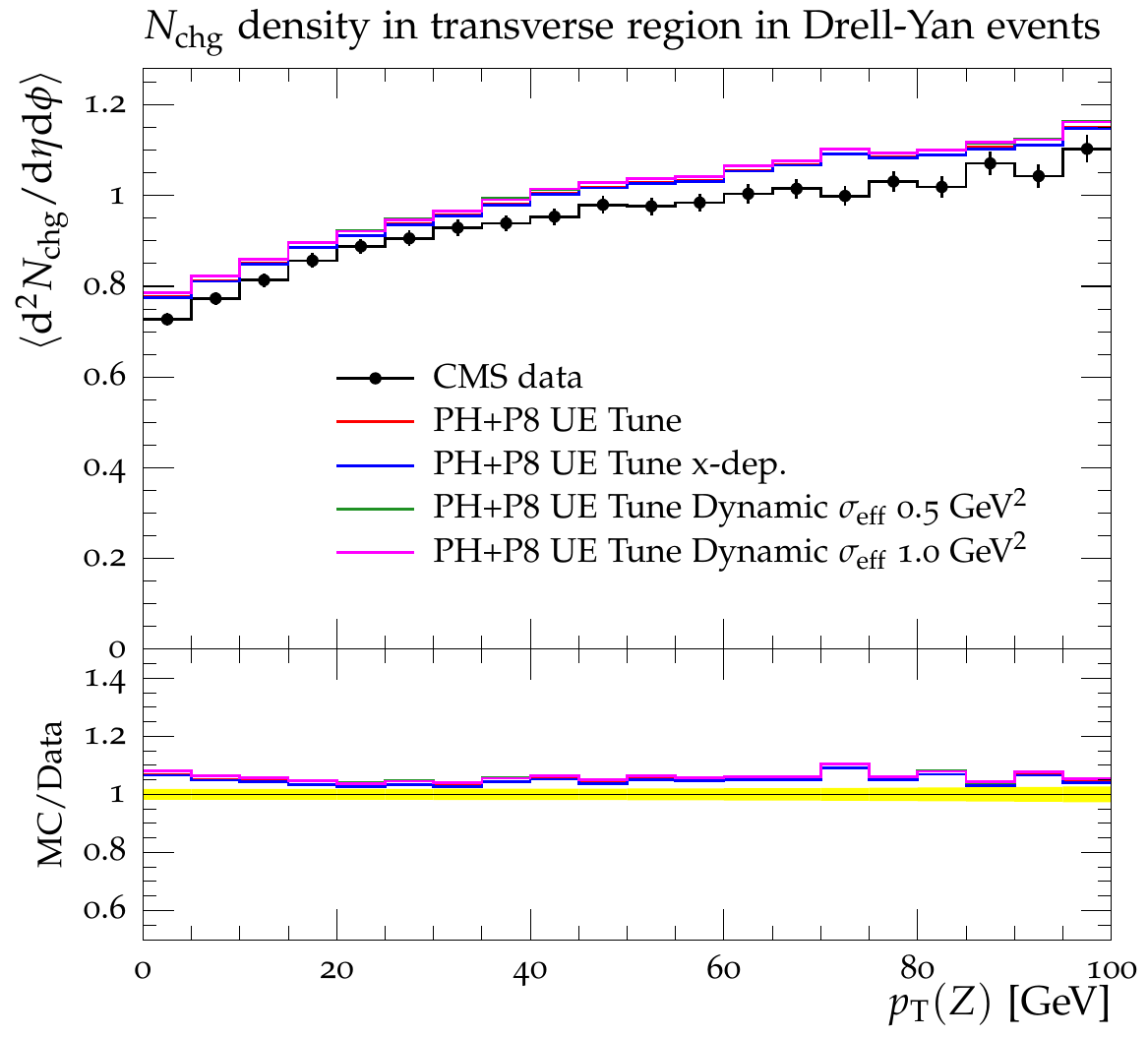}
\includegraphics[scale=0.6]{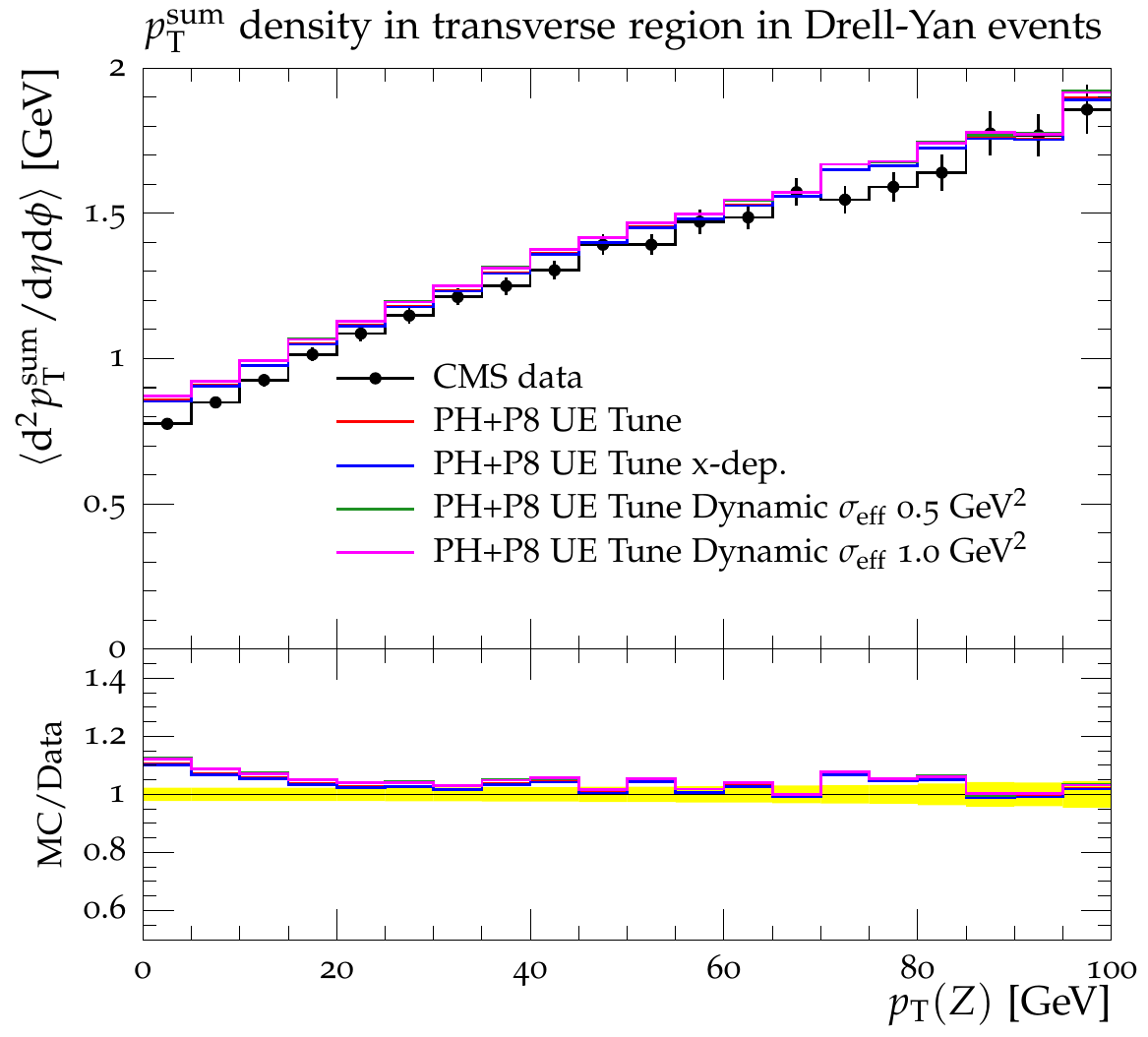}\\
\caption{CMS data~\cite{Chatrchyan:2011id} for the (\textit{left}) charged-particle and (\textit{right}) $p_{\textrm {\scriptsize T}}$ sum densities in the transverse region as defined by the Z-boson in Drell--Yan production in proton-proton collisions at 7 TeV. The data are compared to \textsc{powheg} interfaced to \textsc{pythia} 8 UE Tune with different \effs\ dependence applied: no reweighting applied (red line), $x$-dependent \effs\ values (blue line), $x$- and scale-dependent \effs\ values with $Q^2_0$ $=$ 0.5 GeV$^2$ (green line) and $x$- and scale-dependent \effs\ values with $Q^2_0$ $=$ 1 GeV$^2$ (pink line). Also shown are the ratios of these tunes to the data.}
\label{fig12}
\end{center}
\end{figure}

 In Fig.~\ref{fig12}, various predictions obtained with \textsc{powheg} interfaced to \textsc{pythia}~8 UE Tune are shown of the two UE observables in the transverse region  as a function of the Z-boson $p_{\textrm {\scriptsize T}}$ and compared to the measurement performed by the CMS experiment~\cite{Chatrchyan:2011id}. As seen in inclusive W events, the difference among the considered curves is of the order of 2\%. All predictions are able to follow the data points reasonably well at all scales with differences up to 10\%.

In conclusion, introducing the contribution of \12 mechanisms in the simulation improves the description of measurements of DPS-sensitive observables in Wjj final states. No significant change is observed for variables sensitive to the contribution of moderate MPI and predictions with or without dynamic \effs\ values are able to reproduce the data at the same good level of agreement.

\section{Predictions of DPS-sensitive observables at 14 TeV}

\begin{figure}[htbp]
\begin{center}
\includegraphics[scale=0.67]{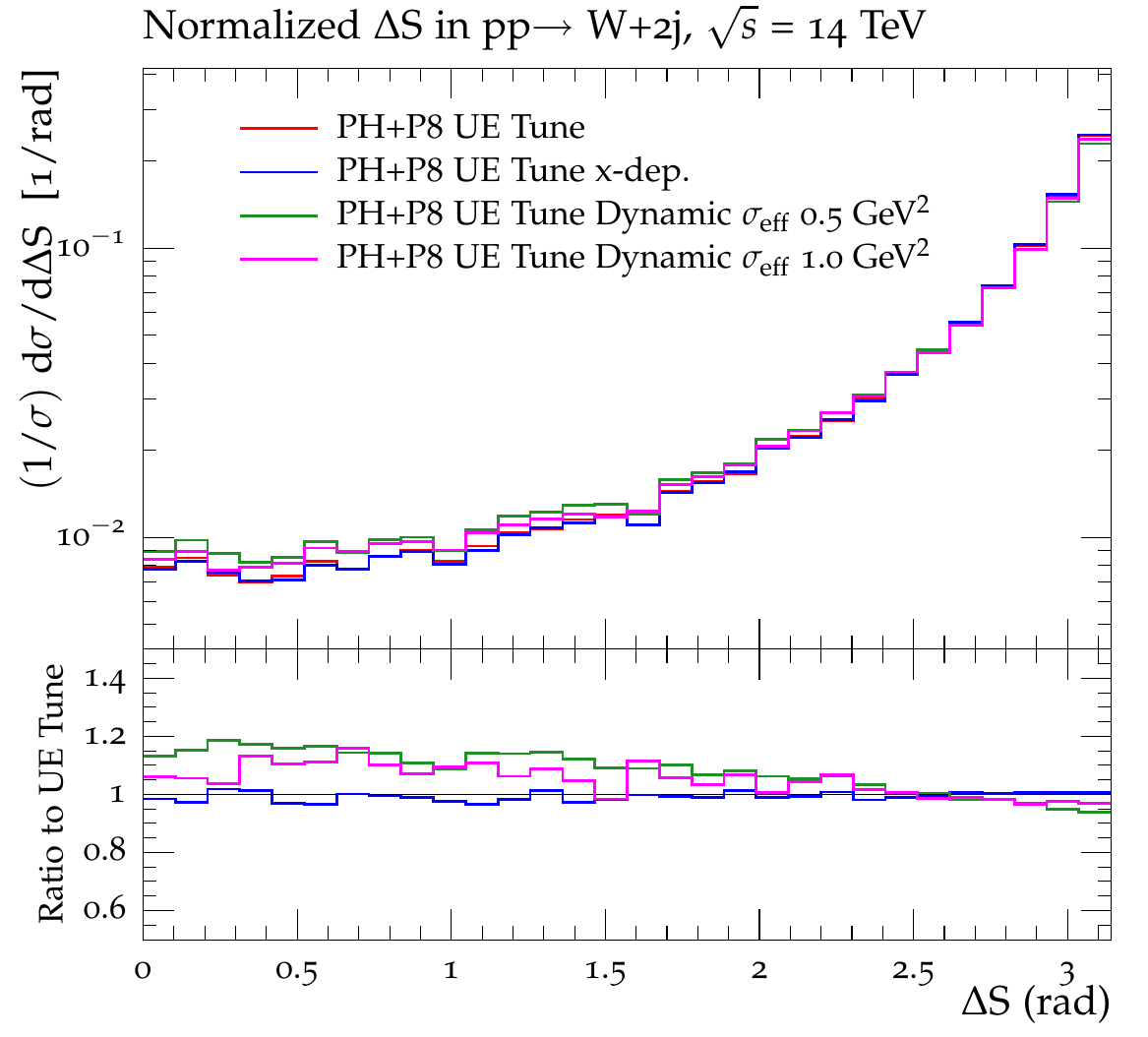}
\includegraphics[scale=0.67]{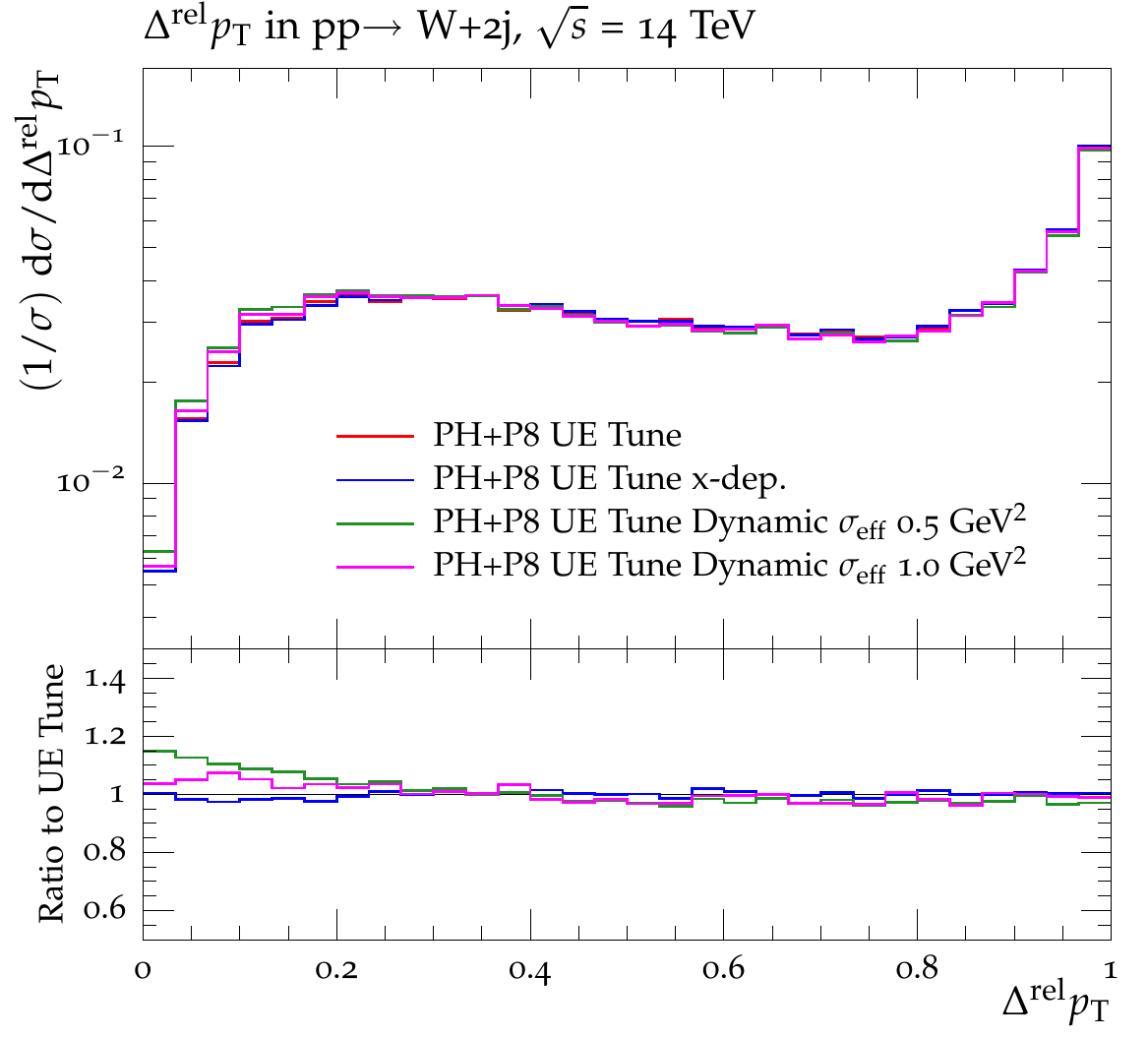}\\
\includegraphics[scale=0.67]{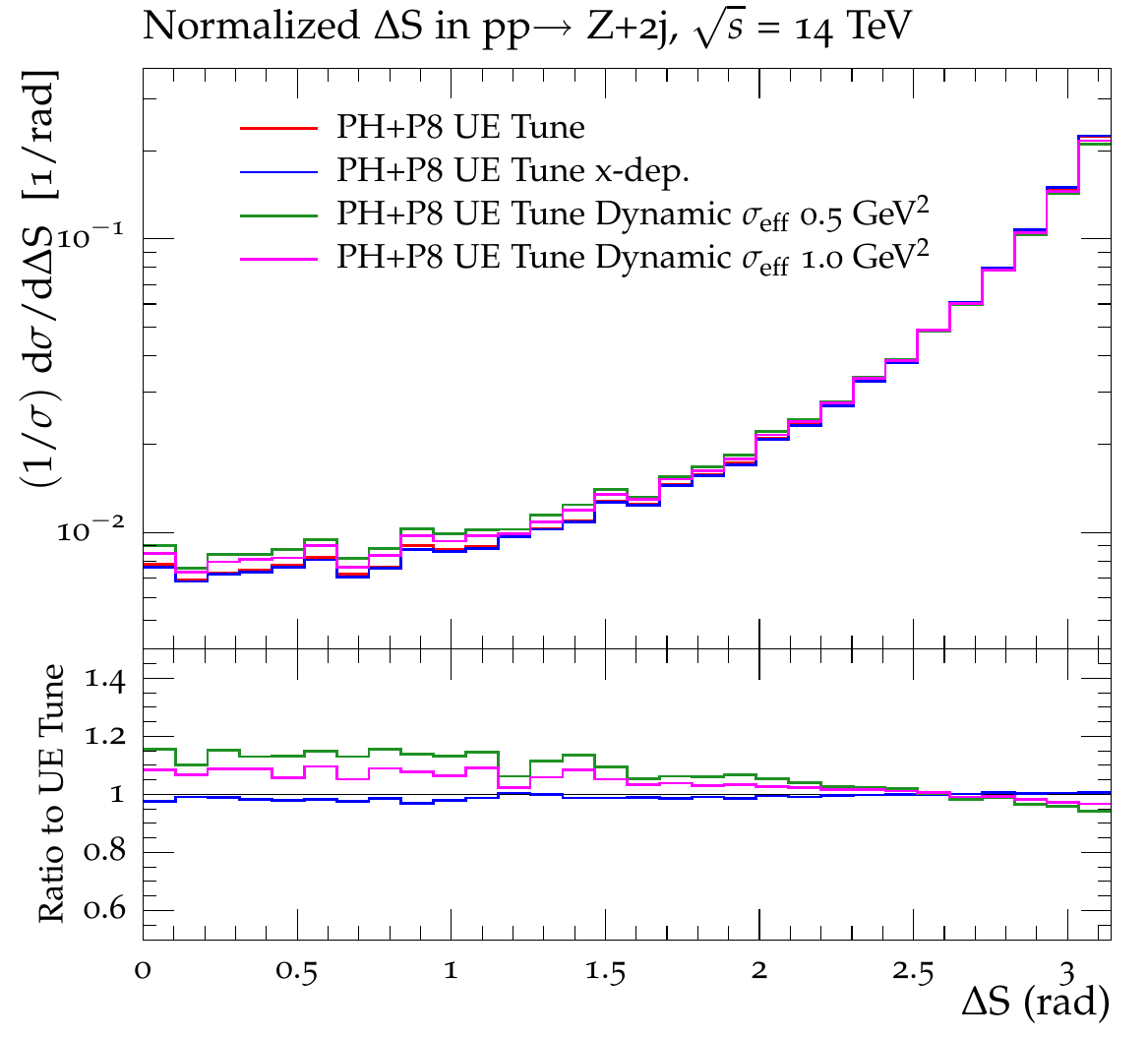}
\includegraphics[scale=0.67]{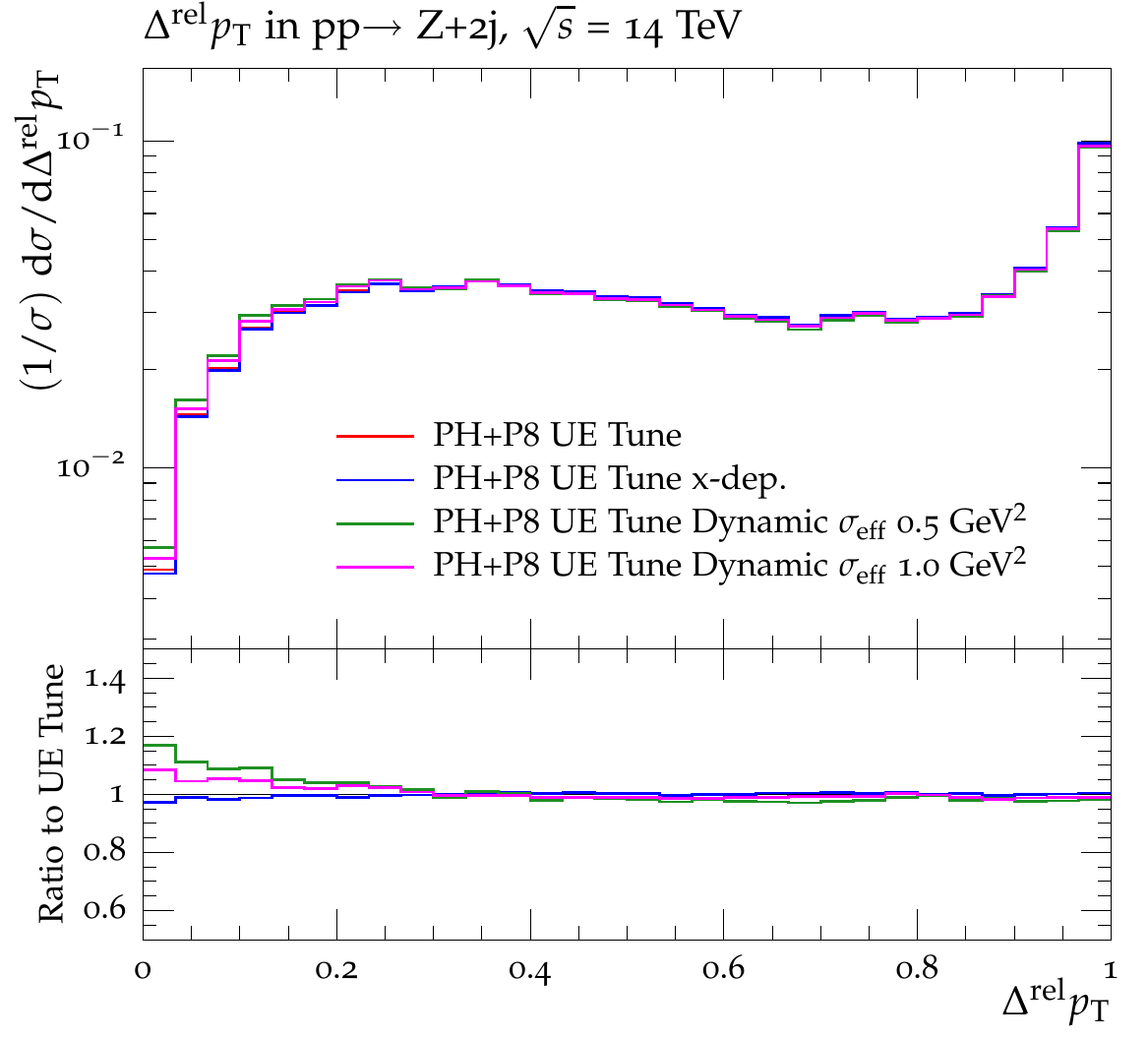}\\
\caption{Predictions at 14 TeV for the normalized distributions of the correlation observables $\Delta$S (\textit{left}) and $\Delta^{\textrm{\scriptsize rel}}p_{\textrm {\scriptsize T}}$ (\textit{right}) in the W+dijet (\textit{top}) and Z+dijet \textit{bottom} channels, of simulations performed with \textsc{powheg} interfaced to \textsc{pythia}~8 UE Tune with different \effs\ dependence applied: no \effs\ reweighting applied (red line), $x$-dependent \effs\ values (blue line), $x$- and scale-dependent \effs\ values with $Q^2_0$ $=$ 0.5 GeV$^2$ (green line) and $x$- and scale-dependent \effs\ values with $Q^2_0$ $=$ 1 GeV$^2$ (pink line). Also shown are the ratios of these tunes to the predictions of the UE Tune.}
\label{fig10}
\end{center}
\end{figure}

In this Section, predictions of DPS-sensitive observables at 14 TeV are shown for Wjj and Zjj final states. Only the \textsc{powheg} event generator is considered with the same settings used for comparisons at 7 TeV. The energy extrapolation of the $p_{\textrm {\scriptsize T}}^0$ value at 14 TeV is applied through the parameter of the \textsc{pythia}~8 tune 4C~\cite{Corke:2010yf}. In \ref{fig10}, predictions at 14 TeV are shown for the normalized distributions of the correlation observables $\Delta$S and $\Delta^{\textrm{\scriptsize rel}}p_{\textrm {\scriptsize T}}$ in the Wjj and Zjj channel, obtained with \textsc{powheg} interfaced to \textsc{pythia}~8 UE Tune with different \effs\ dependence applied: no \effs\ reweighting, $x$-dependent \effs\ values, and  $x$- and scale-dependent \effs\ values with $Q^2_0$ $=$ 0.5 GeV$^2$  and  $Q^2_0$ $=$ 1 GeV$^2$. Very similar behaviour is observed for the two channels. Predictions obtained without any rescaling differ of about 10-15\% from the curves which include the x and scale dependence of \effs\ in the regions of phase space where a DPS signal is expected to contribute, namely $\Delta$S $<$ 2 and $\Delta^{\textrm{\scriptsize rel}}p_{\textrm {\scriptsize T}}$ $<$ 0.2. No relevant difference is observed in case a value of Q$^2_0$ equal to 0.5 or 1.0 GeV$^2$ is used. A higher DPS sensitivity might result for a different jet selection. For instance, bigger differences by about 20--25\% are observed between predictions obtained with \textsc{powheg} interfaced to\textsc{pythia}~8 with and without event reweighting, in case the two jets are selected with a rapidity separation $\Delta\eta$ $>$ 6, in association with a W or a Z boson. A requirement of a large $|\Delta\eta|$ indeed suppresses the contribution of SPS processes and increases the sensitivity to DPS contributions.

\newpage

\section{Conclusions}
\par The new tune, developed in \cite{BG} for the analysis of multiple parton interactions (MPI) in four-jet final states and including contributions from \12 mechanisms, is compatible with measurements sensitive to double parton scattering (DPS) and MPI at moderate scales in W+dijet and Z+dijet channels. In order to properly treat events with a W or a Z boson with associated jets, it is necessary to include higher-order contributions within the matrix element calculation. In this paper, simulations using the \textsc{powheg} event generator interfaced to the underlying event (UE) simulation provided by \textsc{pythia}~8 are considered. Predictions using dynamic \effs\ values dependent on the longitudinal momentum fractions and on the scale of the process improve the description of correlation observables measured in W+dijet final states. The experimental accuracy achieved so far does not allow to make conclusion on the best value of the scale separating soft and hard MPI, Q$^2_0$. Values of Q$^2_0$ between 0.5 and 1 are able to describe the measurement at the same level of agreement. Differences by about 15\% are observed between these predictions and measured data if the jet balance in transverse momentum $p_{\textrm {\scriptsize T}}$ ($\Delta^{\textrm{\scriptsize rel}}p_{\textrm {\scriptsize T}}$) is considered. This might be due to the assumption made in our simulation that differential distributions in \12 mechanisms have the same $p_{\textrm {\scriptsize T}}$ dependence as the conventional \22 diagram, as a simple generalization of the formula given in~\cite{BDFS2,BDFS3,BDFS4}. Although explicit formulae for differential distributions of \12 production mechanisms are known \cite{BDFS3,BDFS4}, their actual implementation still demands additional work, both analytical and numerical.

\section*{Acknowledgements}
We thank M. Strikman and H. Jung for very useful discussions and reading the manuscript.

\appendix
\section{Results from predictions using MADGRAPH}

In this section, we consider results obtained by considering the \textsc{madgraph} event generator. This is interesting to evaluate the contribution of virtual NLO corrections which are included in \textsc{powheg}, but not in the calculation of the matrix element implemented in \textsc{madgraph}. Furthermore, the \textsc{madgraph} event generator has been used as reference for the extraction of the DPS contribution in experimental measurements~\cite{Atlas,cms1,cms2,Chatrchyan:2013qza}.\\

\begin{figure}[htbp]
\begin{center}
\includegraphics[scale=0.67]{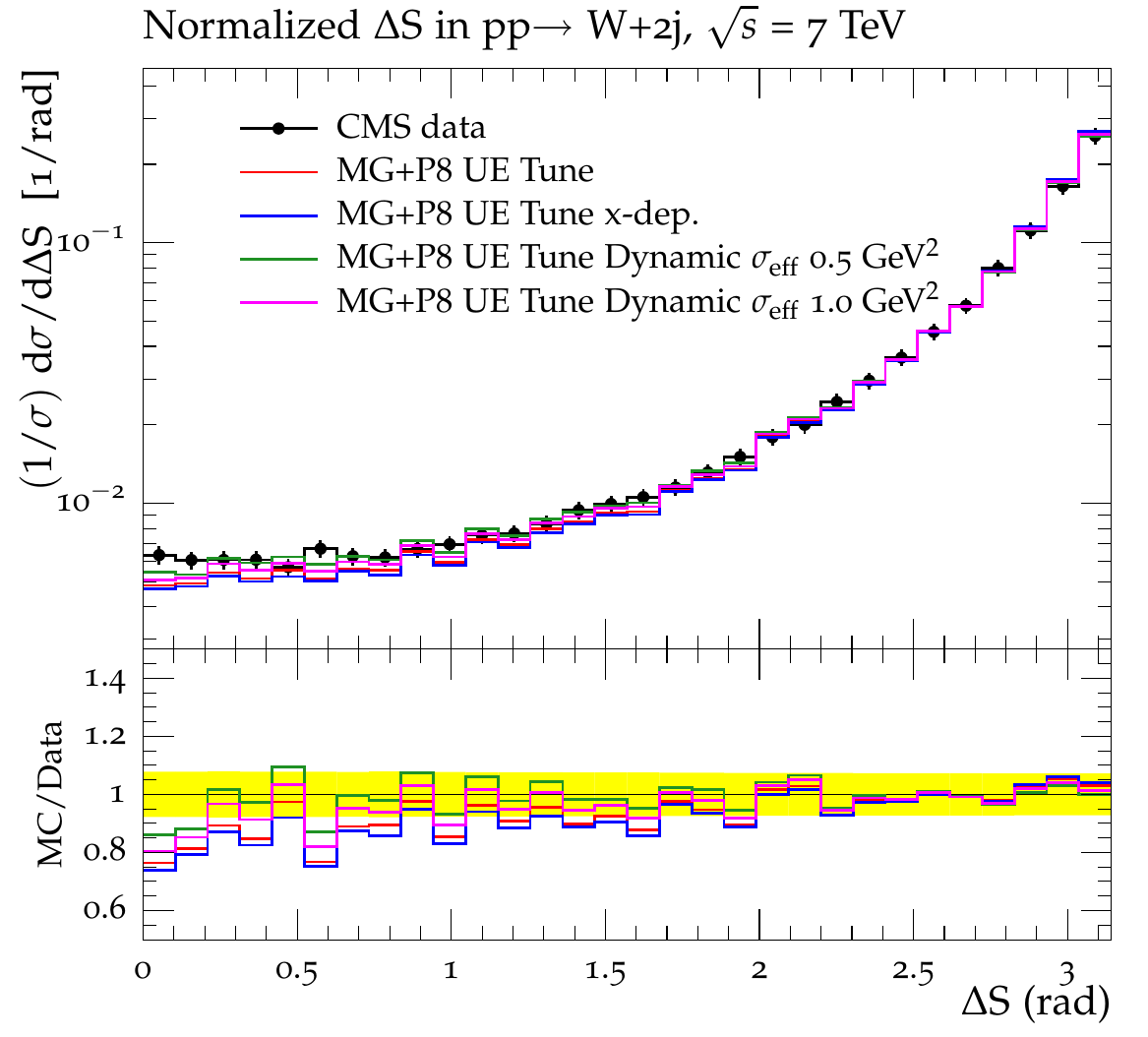}
\includegraphics[scale=0.67]{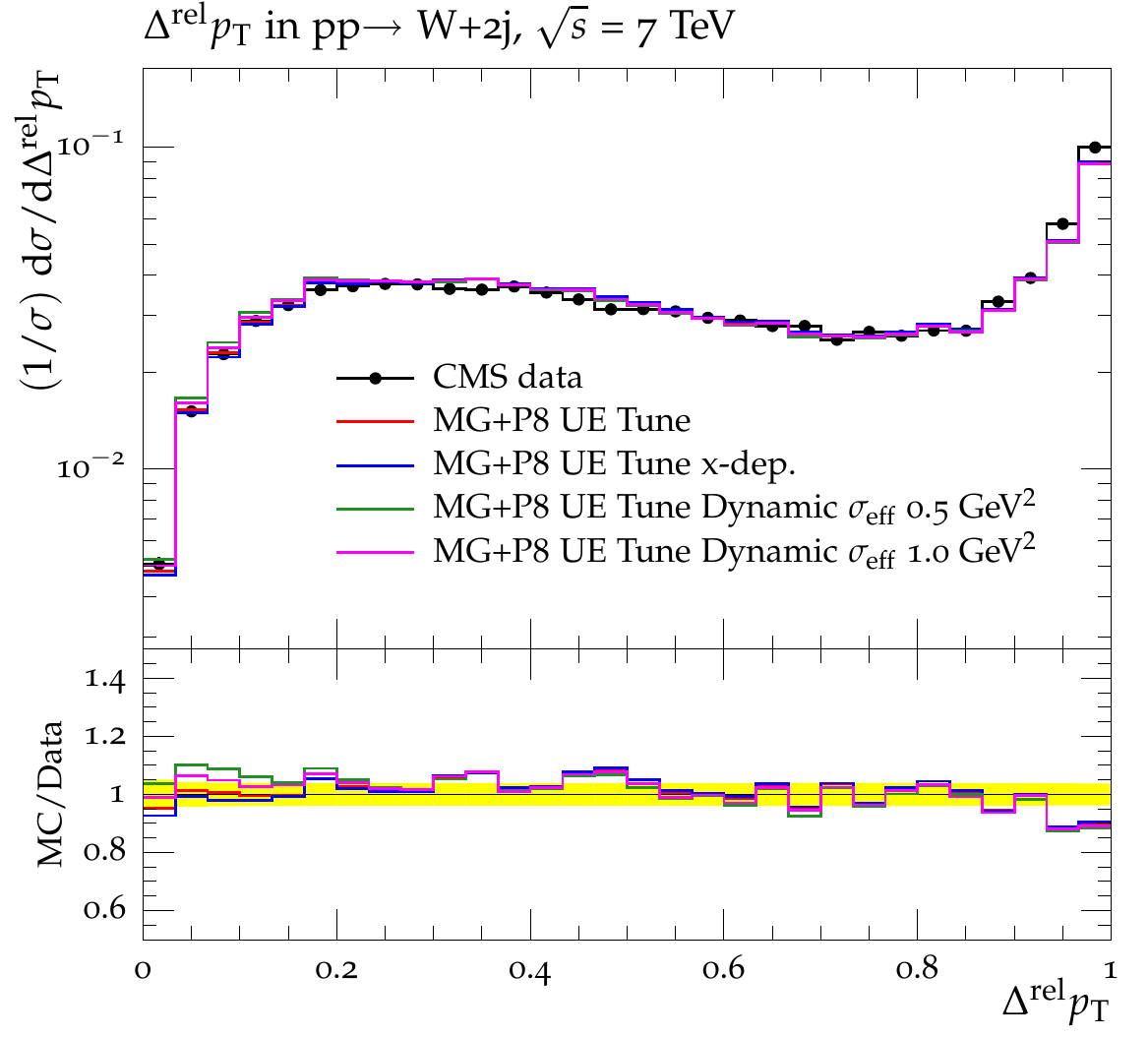}\\
\includegraphics[scale=0.67]{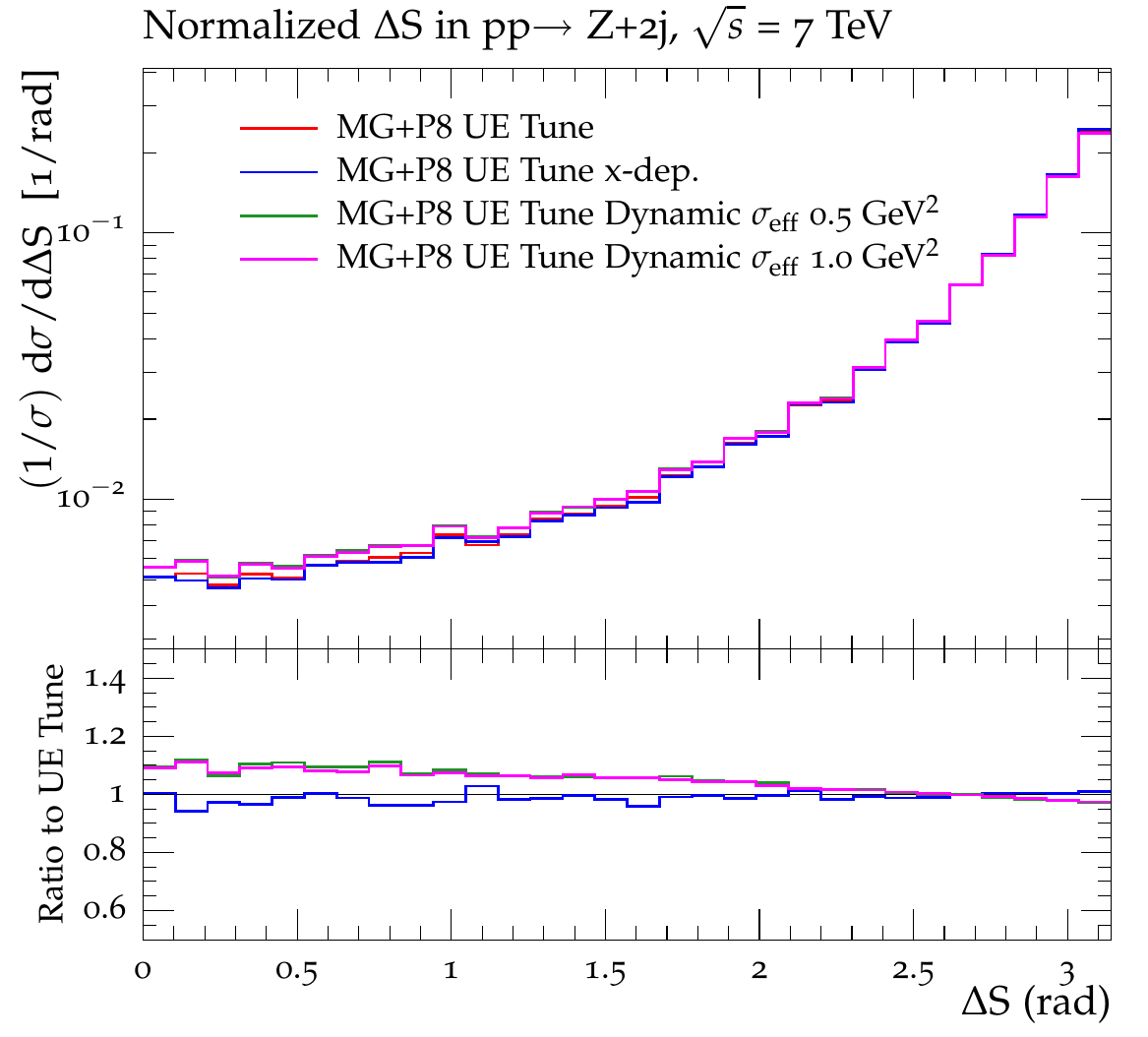}
\includegraphics[scale=0.67]{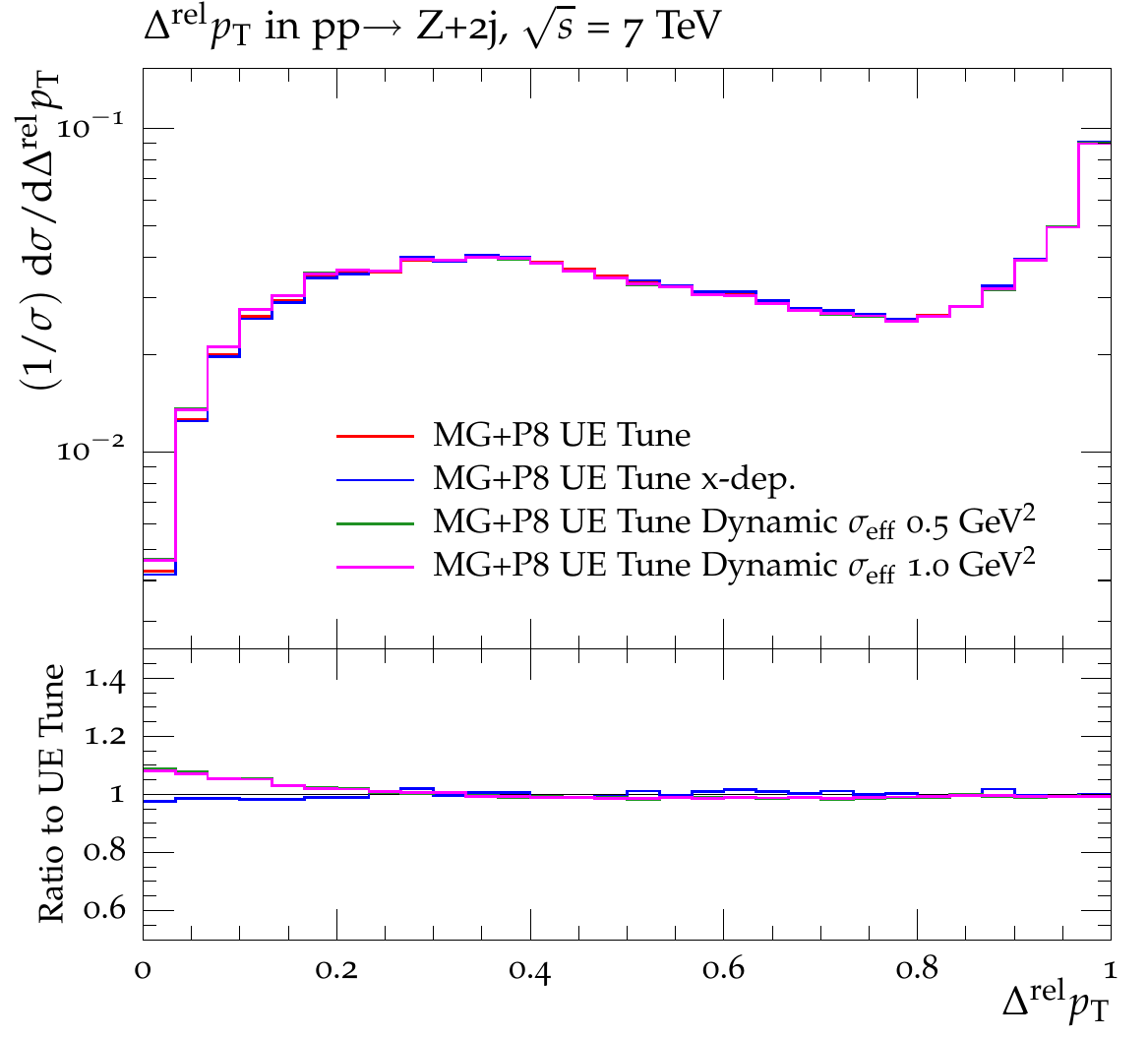}\\
\caption{\textit{top} CMS data at 7 TeV for the normalized distributions of the correlation observables $\Delta$S (\textit{left}) and $\Delta^{\textrm{\scriptsize rel}}p_{\textrm {\scriptsize T}}$ (\textit{right}) in the W+dijet channel, compared to predictions of \textsc{madgraph} interfaced to \textsc{pythia}~8 UE Tune with different \effs\ dependence applied: no reweighting applied (red line), $x$-dependent \effs\ values (blue line), $x$- and scale-dependent \effs\ values with $Q^2_0$ $=$ 0.5 GeV$^2$ (green line) and $x$- and scale-dependent \effs\ values with $Q^2_0$ $=$ 1 GeV$^2$ (pink line). Also shown are the ratios of these predictions to the data. \textit{bottom} Predictions at 7 TeV for the normalized distributions of the correlation observables $\Delta$S (\textit{left}) and $\Delta^{\textrm{\scriptsize rel}}p_{\textrm {\scriptsize T}}$ (\textit{right}) in the Z+dijet channel, of simulations performed with \textsc{madgraph} interfaced to \textsc{pythia}~8 UE Tune with different \effs\ dependence applied: no \effs\ reweighting applied (red line), $x$-dependent \effs\ values (blue line), $x$- and scale-dependent \effs\ values with $Q^2_0$ $=$ 0.5 GeV$^2$ (green line) and $x$- and scale-dependent \effs\ values with $Q^2_0$ $=$ 1 GeV$^2$ (pink line). Also shown are the ratios of these tunes to the predictions of the UE Tune.}
\label{fig2}
\end{center}
\end{figure}

In Fig.~\ref{fig2}(top), measurements from the CMS experiment at 7 TeV of normalized cross sections as a function of $\Delta$S and $\Delta^{\textrm{\scriptsize rel}}p_{\textrm {\scriptsize T}}$ in the Wjj channel are compared to predictions obtained with \textsc{madgraph} interfaced to \textsc{pythia}~8 using various \effs\ settings. Predictions with a constant value of \effs, with x dependence applied and with x and scale dependence applied with $Q^2_0$ $=$ 0.5 and 1.0 GeV$^2$ are investigated. Predictions with dynamic \effs\ values dependent on x and on the scale describe better the measurement, especially at values of $\Delta$S $<$ 2 and at $\Delta^{\textrm{\scriptsize rel}}p_{\textrm {\scriptsize T}}$ $<$ 0.2. A slight better agreement than predictions obtained with the \textsc{powheg} matrix element is observed for the normalized cross section as a function of $\Delta^{\textrm{\scriptsize rel}}p_{\textrm {\scriptsize T}}$. As for the \textsc{powheg} case, it is not possible to discriminate the best value of $Q^2_0$, due to the large experimental uncertainty. In Fig.~\ref{fig2}(bottom), predictions with the same settings are tested on $\Delta$S and $\Delta^{\textrm{\scriptsize rel}}p_{\textrm {\scriptsize T}}$ in the Zjj channel. As the results for \textsc{powheg}, differences between predictions with dynamic \effs\ values are 10-15\% above the curve using no \effs\ rescaling at $\Delta$S $<$ 2 and at $\Delta^{\textrm{\scriptsize rel}}p_{\textrm {\scriptsize T}}$ $<$ 0.2.

\begin{figure}[htbp]
\begin{center}
\includegraphics[scale=0.6]{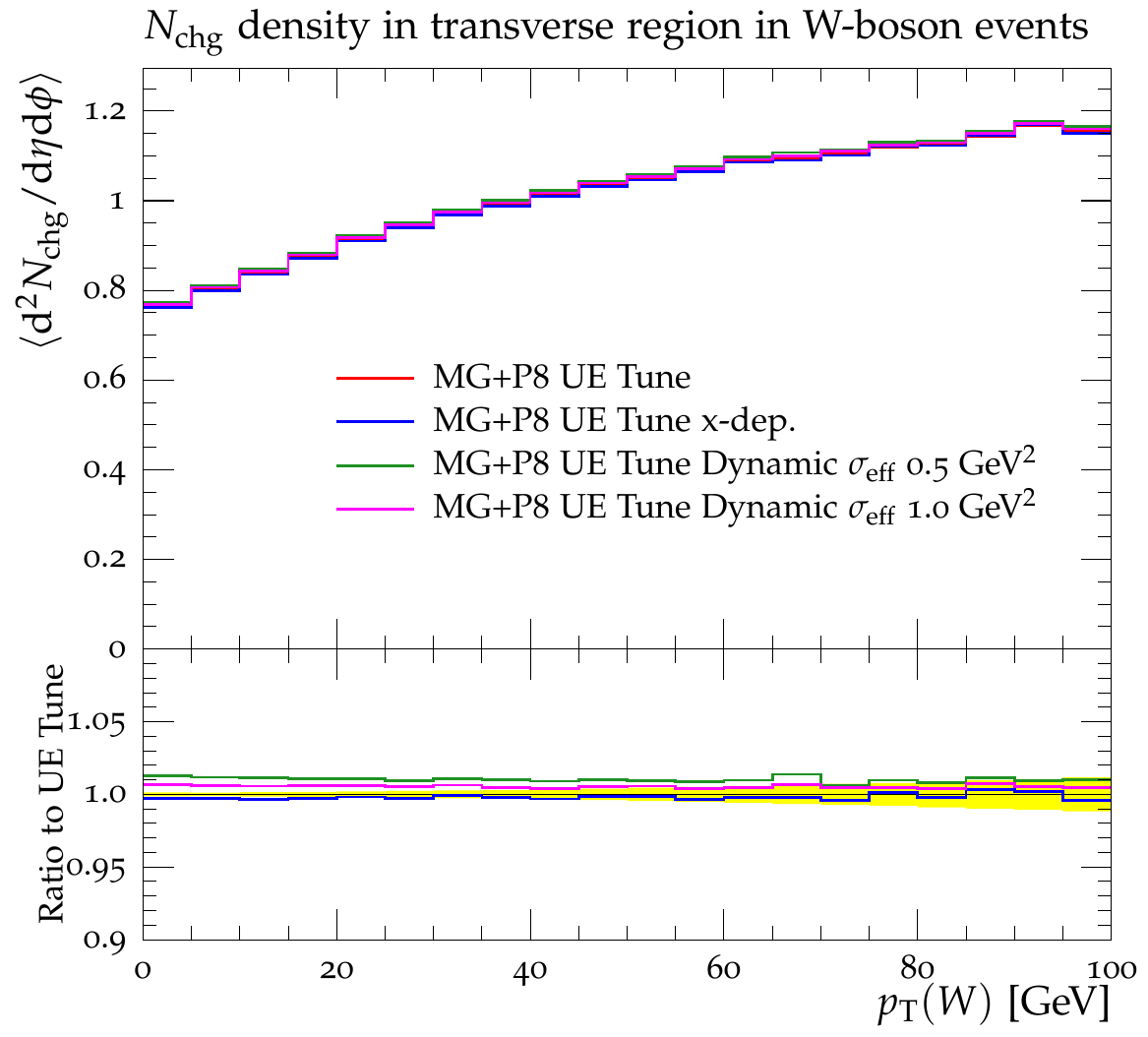}
\includegraphics[scale=0.6]{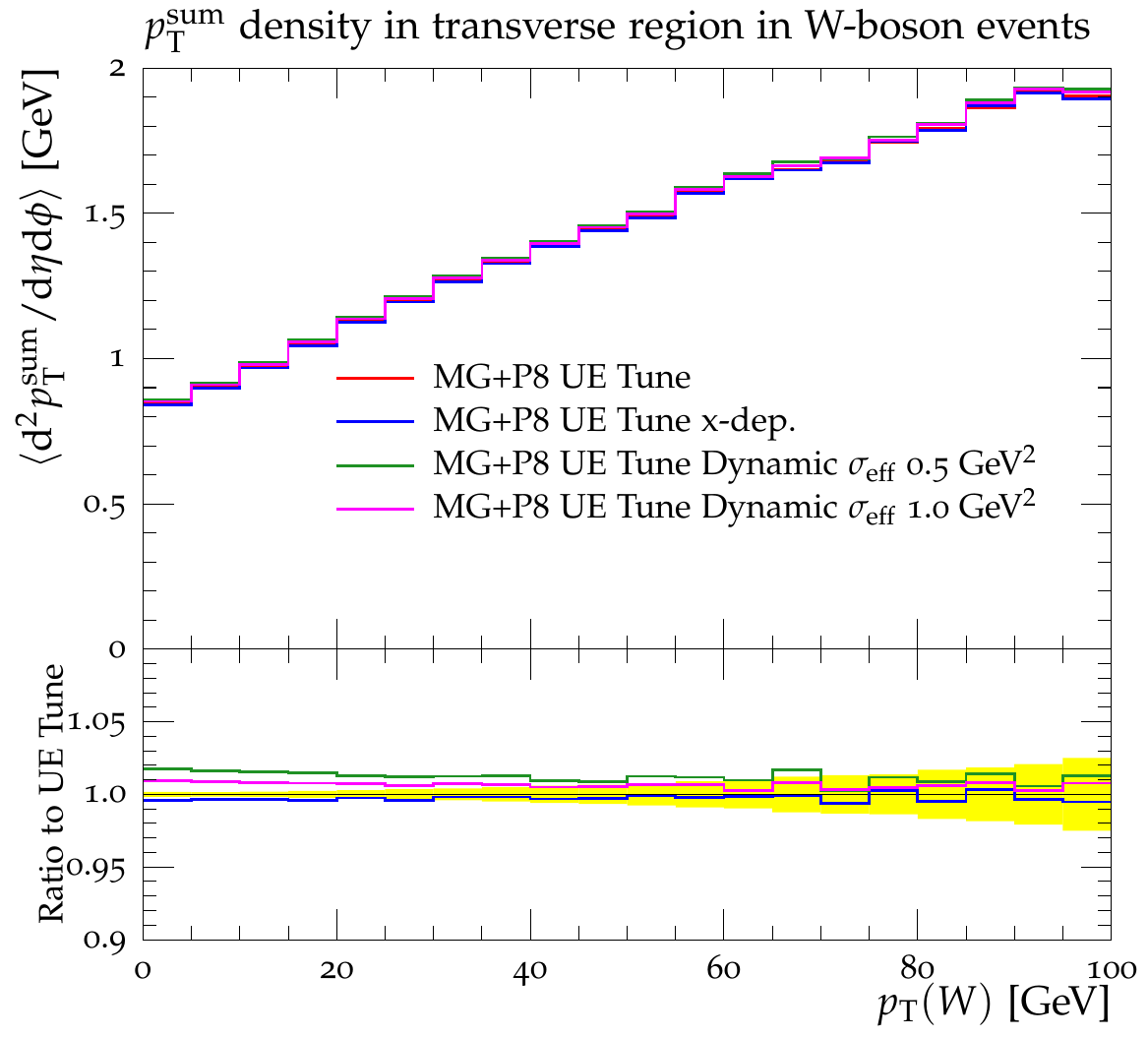}\\
\includegraphics[scale=0.6]{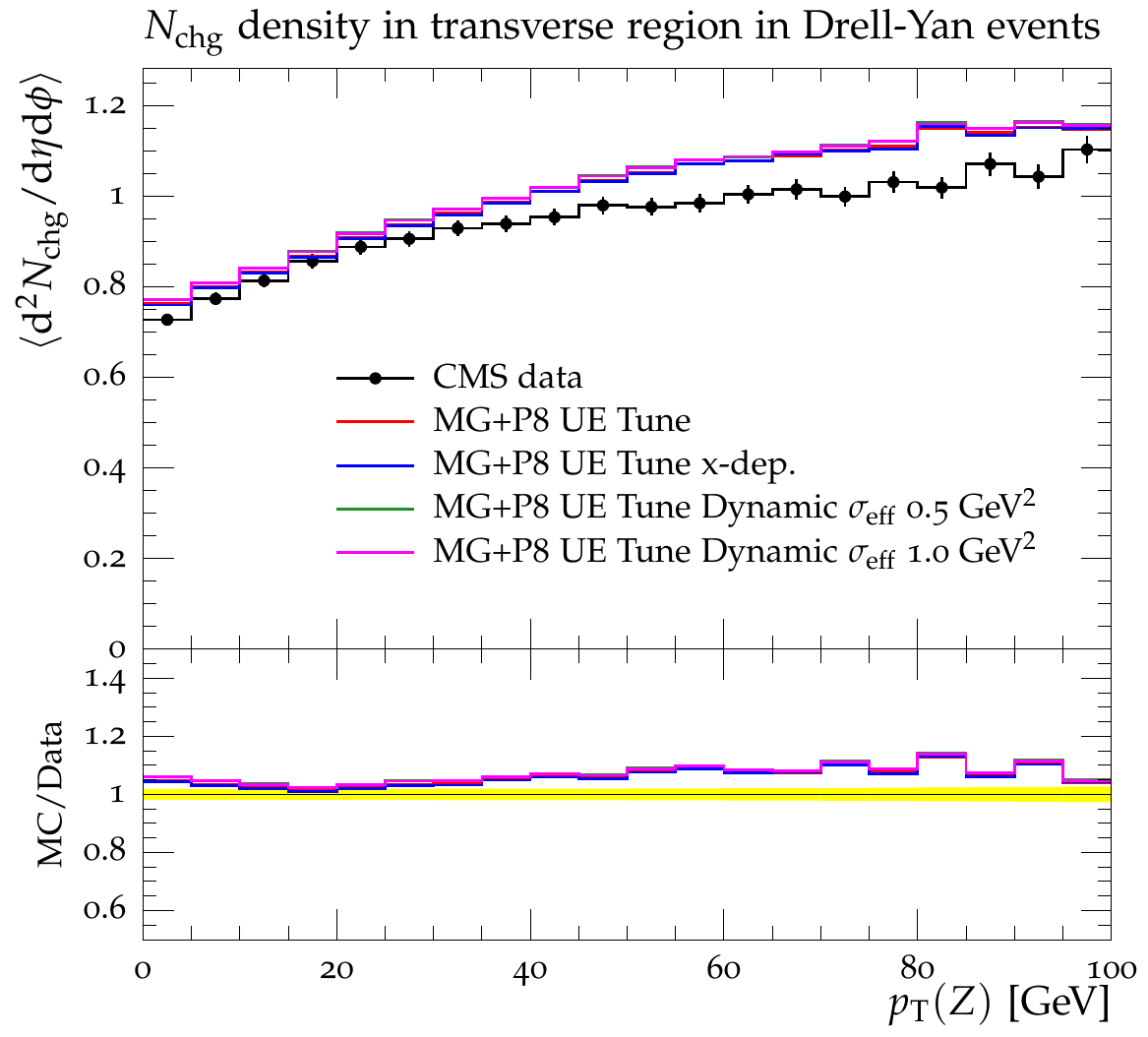}
\includegraphics[scale=0.6]{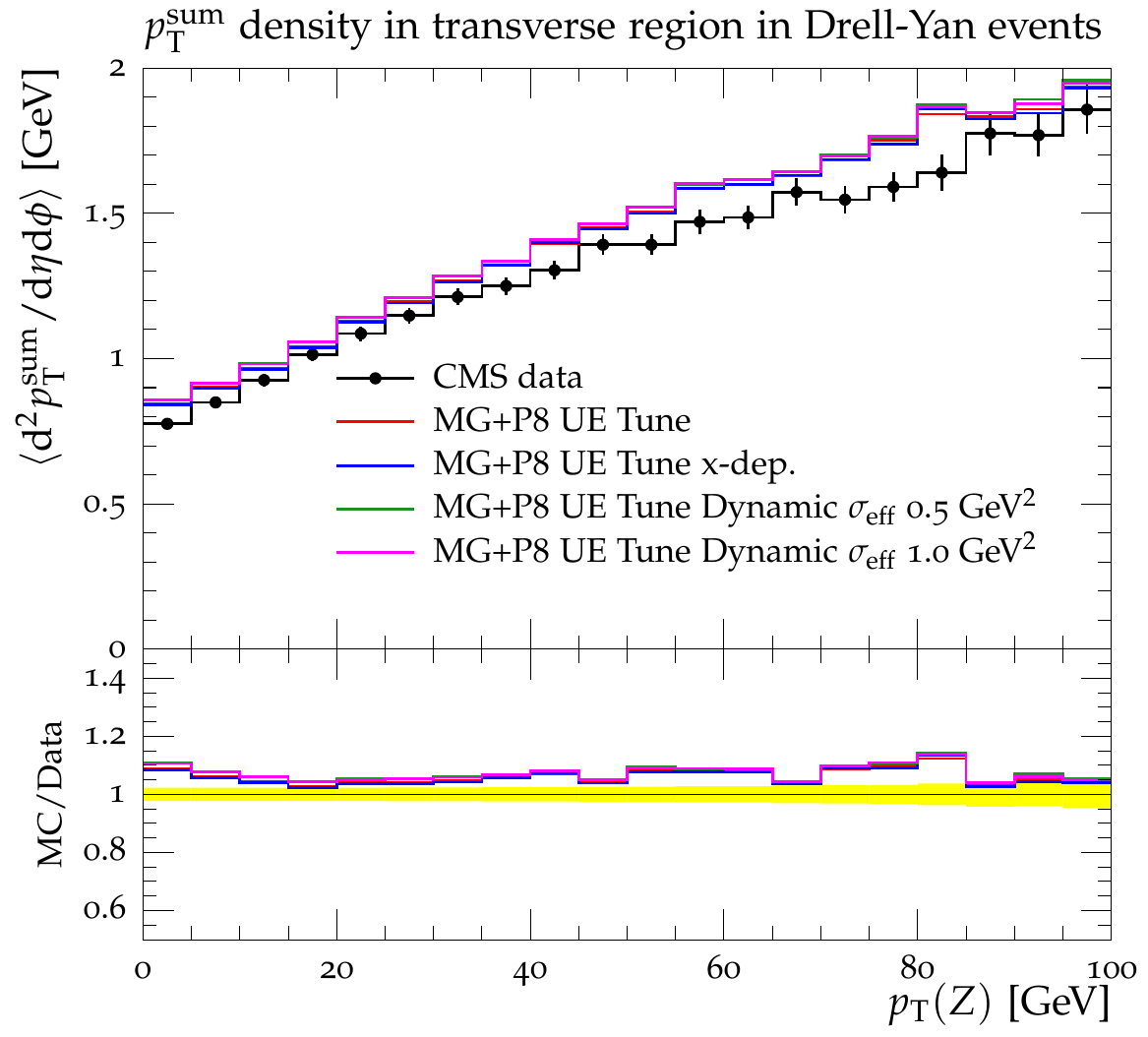}
\caption{(\textit{top}) Predictions for the (\textit{left}) charged-particle and (\textit{right}) $p_{\textrm {\scriptsize T}}$ sum densities in the transverse regions as defined by the W-boson in proton-proton collisions at 7 TeV. Simulations obtained with \textsc{madgraph} interfaced to \textsc{pythia}~8 UE Tune are considered with different \effs\ dependence applied: no reweighting applied (red line), $x$-dependent \effs\ values (blue line), $x$- and scale-dependent \effs\ values with $Q^2_0$ $=$ 0.5 GeV$^2$ (green line) and $x$- and scale-dependent \effs\ values with $Q^2_0$ $=$ 1 GeV$^2$ (pink line). Also shown are the ratios of these tunes to the predictions of the UE Tune. (\textit{bottom}) CMS data for the (\textit{left}) charged-particle and (\textit{right}) $p_{\textrm {\scriptsize T}}$ sum densities in the transverse region as defined by the Z-boson in Drell--Yan production in proton-proton collisions at 7 TeV. The data are compared to \textsc{madgraph} interfaced to \textsc{pythia} 8 UE Tune with different \effs\ dependence applied: no reweighting applied (red line), $x$-dependent \effs\ values (blue line), $x$- and scale-dependent \effs\ values with $Q^2_0$ $=$ 0.5 GeV$^2$ (green line) and $x$- and scale-dependent \effs\ values with $Q^2_0$ $=$ 1 GeV$^2$ (pink line). Also shown are the ratios of these predictions to the data.}
\label{fig5}
\end{center}
\end{figure}

In Fig.~\ref{fig5}, the observables sensitive to the UE are investigated by comparing different \textsc{madgraph} predictions. In Fig.~\ref{fig5}(top), predictions on the charged particle multiplicity and $p_{\textrm {\scriptsize T}}$ sum densities as a function of the W-boson $p_{\textrm {\scriptsize T}}$ are shown in the inclusive W channel, while in Fig.~\ref{fig5}(bottom) the same predictions are compared to the CMS measurement in the inclusive Z channel. Very similar conclusions as drawn when considering \textsc{powheg} can be extracted. Differences among the various predictions are observed only of the order of 2\% and, in the case of inclusive Z production, all of them are able to reproduce the trend of the measured points.\\

In conclusion, DPS-sensitive observables are better described by predictions using dynamic \effs\ values, while variables sensitive to MPI at moderate scales are not strongly affected by \effs\ variation within our approach. This is also the case for \textsc{powheg}. Considering the fact that same conclusions hold for \textsc{madgraph} and \textsc{powheg}, real NLO corrections are dominant in Wjj and Zjj final states, while virtual ones have a low impact.

\clearpage
\newpage

\section{Values of sigma effective}

\par
\begin{figure}[htbp]
\begin{center}
\includegraphics[scale=0.37]{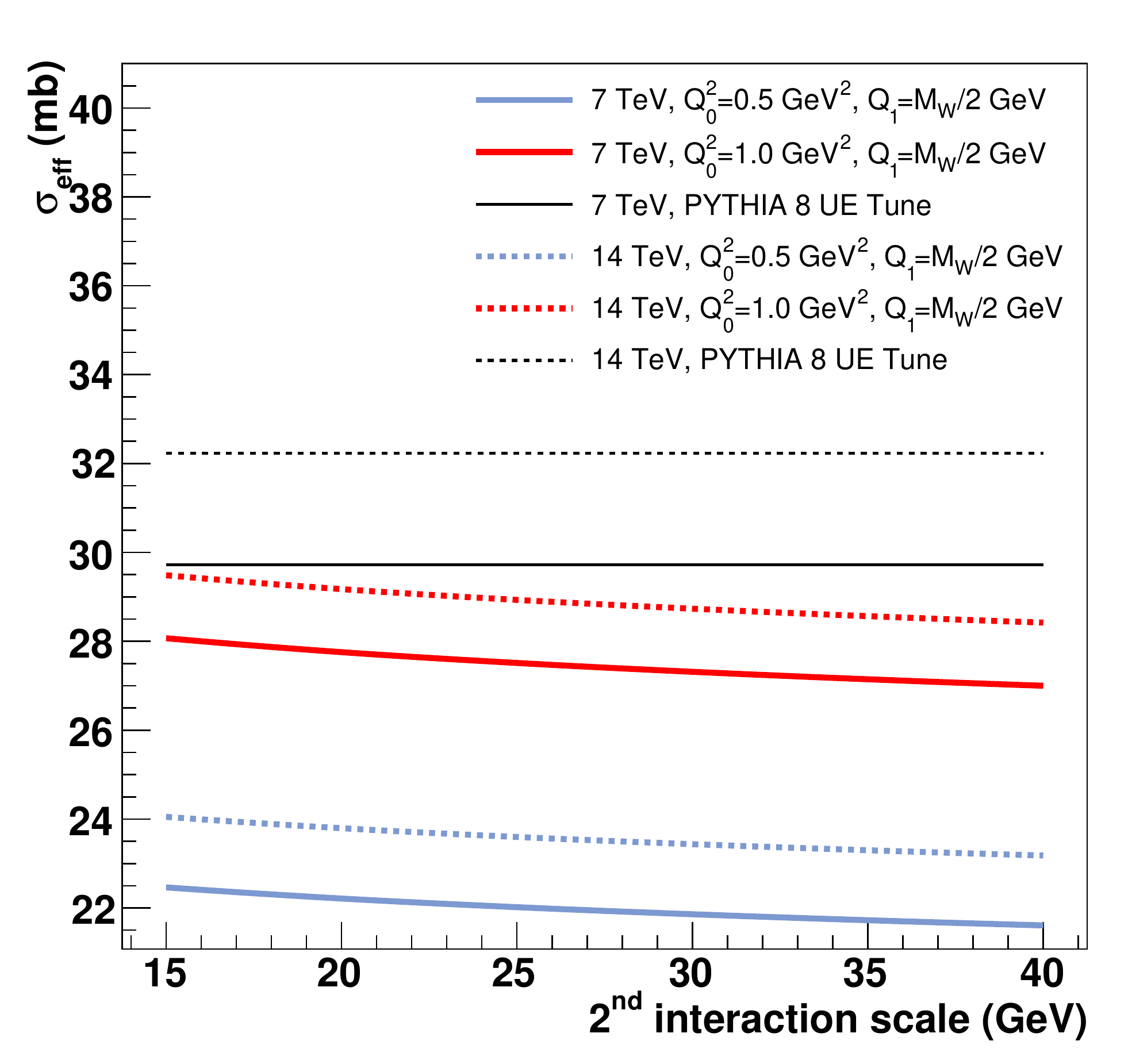}
\includegraphics[scale=0.37]{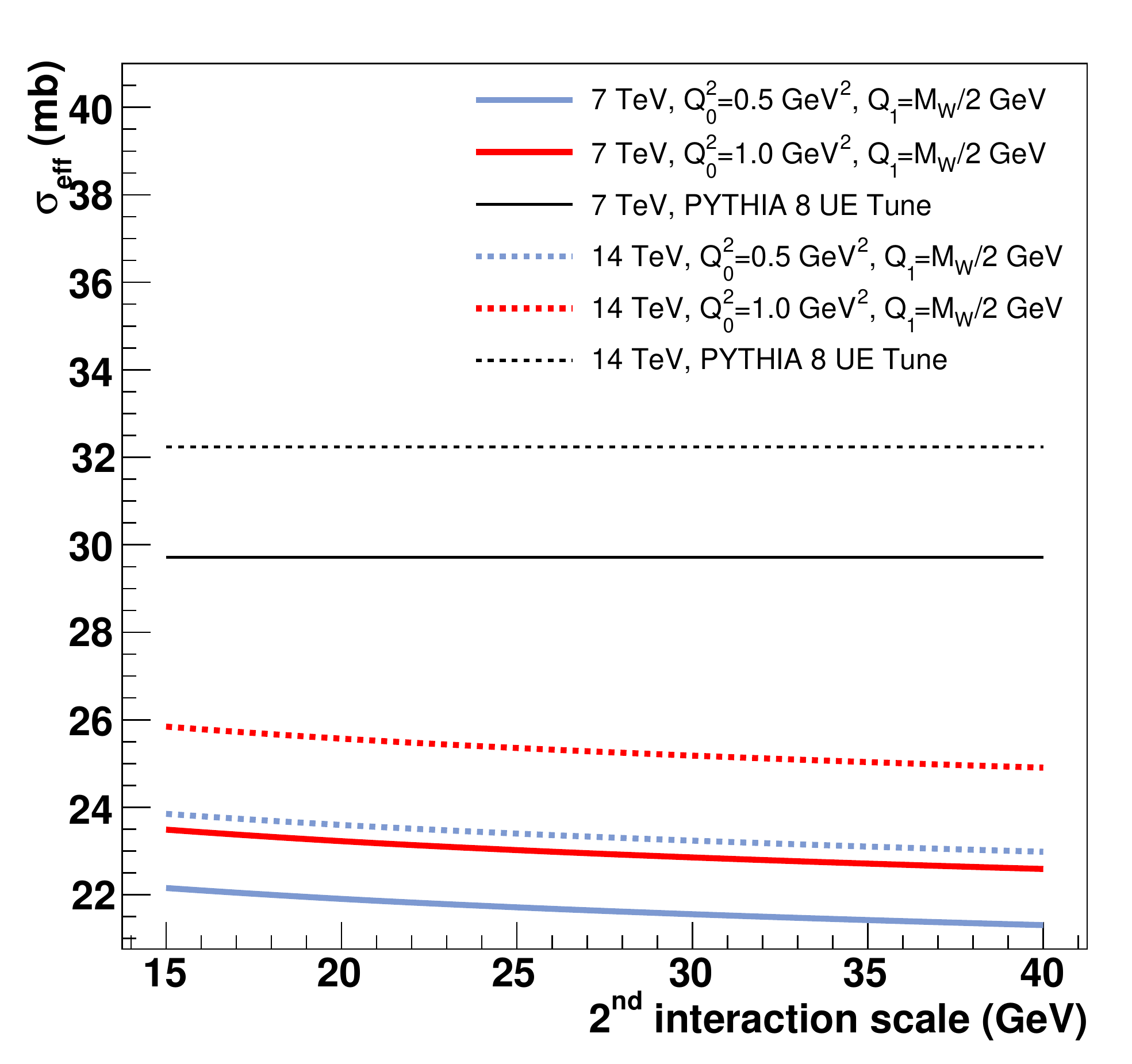}
\caption{Values of \effs\ as a function of the scale of the 2$^{nd}$ interaction at different collision energies at 7 TeV and 14 TeV for first hard interactions occurring at a scale $Q_1$ equal to M$_W$/2 and M$_Z$/2 GeV for, respectively, Wjj (\textit{left}) and Zjj (\textit{right}) channels. The two values of $Q_0^2$ equal to 0.5 and 1.0 GeV$^2$ are considered and the longitudinal momentum fractions of the two dijets correspond to the maximal transverse momentum exchange for both $\sqrt{s}$ = 7 TeV and $\sqrt{s}$ = 14 TeV. Also shown are the values of \effs\ for each energy, as implemented in the \textsc{pythia}~8 UE Tune if no reweighting is applied.}
\label{figApp1}
\end{center}
\end{figure}

 In this section, a closer look at the \effs\ dependence as a function of collision energy and parton scales is taken. Fig.~\ref{figApp1} shows the values of \effs\ as a function of the scale of the secondary hard scattering for values of $Q^2_0$ of 0.5 and 1.0 GeV$^2$ at $\sqrt{s}$ = 7 and 14 TeV for the considered channels, Wjj and Zjj. The scale of the hard scattering is kept fixed to the maximum of the Breit-Wigner distribution, namely M$_W$/2 and M$_Z$/2 for, respectively, Wjj and Zjj final states. Values of \effs\ are slowly decreasing as a function of the scale of the secondary interaction and change of about 1 mb between 15 and 40 GeV, independently on the Q$^2_0$ value. The change in center-of-mass energy brings the value of \effs\ up of about 3-5 mb. Similar conclusions can be extracted from the two considered channels.

\clearpage

\bibliography{mpi21}

\end{document}